\documentclass{aastex62}

\usepackage{graphicx}
\usepackage{natbib}
\usepackage{subfigure}
\usepackage{epsfig}
\usepackage{siunitx}
\usepackage{url}

\bibpunct{(}{)}{;}{a}{}{,} % to follow the A&A style
\newcommand*{\eepair}  {$\mathrm{e}^+-\mathrm{e}^-$}
\newcommand*{\aeff}  {$A_\mathrm{eff}$}
\newcommand*{\aeffm}  {A_\mathrm{eff}}
\newcommand*{\egamma}  {E_\mathrm{\gamma}}
\newcommand*{\egrid}  {E_\mathrm{GRID}}
\newcommand*{\epts}  {E_\mathrm{PTS}}

\submitjournal{ApJ}

\shorttitle{Calibration of AGILE-GRID on-ground}
\shortauthors{Cattaneo~P.W. et al.}

\begin{document}

\title{Calibration of AGILE-GRID with on-ground data and Monte Carlo simulations}

%\include{author-list-aa}
%\correspondingauthor{P.W. Cattaneo}
%\email{paolo.cattaneo@pv.infn.it}

\newcommand*{\infnpv}  {INFN-Pavia, Via A. Bassi 6, I-27100 Pavia, Italy}
\newcommand*{\inafro}  {INAF/IAPS-Roma, via del Fosso del Cavaliere 100, I-00133 Roma, Italy}
\newcommand*{\infnts}  {INFN-Trieste, Via A. Valerio 2, I-34127 Trieste, Italy} 
\newcommand*{\inafbo}  {INAF/IASF-Bologna, Via P. Gobetti 101, I-40129 Bologna, Italy} 
\newcommand*{\infnlnf} {INFN Lab. Naz. di Frascati, Via E. Fermi, 40, I-00044 Frascati(Roma), Italy} 
\newcommand*{\uniwits} {School of Physics, Wits University, Johannesburg, South Africa} 
\newcommand*{\oar} {INAF-Osservatorio Astronomico di Roma, Via Frascati 33, I-00078 Monte Porzio Catone (Roma), Italy} 
\newcommand*{\inafirabo} {INAF-IRA Bologna, Via Gobetti 101, I-40129 Bologna, Italy} 
\newcommand*{\asi} {Agenzia Spaziale Italiana (ASI), Via del Politecnico snc, I-00133 Roma, Italy} 
\newcommand*{\cifs} {CIFS, c/o Dip. Fisica, Univ. di Torino, Via P. Giuria 1, I-10125 Torino, Italy} 
\newcommand*{\unirotv} {Dip. di Fisica, Univ. Tor Vergata, Via della Ricerca Scientifica 1, I-00133 Roma,Italy} 
\newcommand*{\eneabo} {ENEA-Bologna, Via Martiri di Monte Sole 4, I-40129 Bologna, Italy} 
\newcommand*{\inafmi} {INAF/IASF-Milano, Via E. Bassini 15, I-20133 Milano, Italy} 
\newcommand*{\units} {Dip. Fisica, Univ. di Trieste, Via A. Valerio 2, I-34127 Trieste, Italy} 
\newcommand*{\ssdc} {ASI Space Science Data Center (SSDC), Via del Politecnico snc, I-00133 Roma, Italy} 
\newcommand*{\bergen} {Birkeland Centre for Space Science, Department of Physics and Technology, University of Bergen, Norway} 
\newcommand*{\infnrotv} {INFN-Roma Tor Vergata, Via della Ricerca Scientifica 1, I-00133 Roma, Italy} 
\newcommand*{\ewrsd} {East Windsor RSD, 25a Leshin Lane, Hightstown, NJ 08520, USA} 
\newcommand*{\oac} {INAF-Osservatorio Astronomico di Cagliari, Via della Scienza 5, I-09047 Selargius (CA), Italy}
\newcommand*{\unico} {Univ. dell'Insubria, Via Valleggio 11, I-22100 Como, Italy} 
\newcommand*{\eneafr} {ENEA-Frascati, Via E. Fermi, 45, I-00044 Frascati (Roma), Italy} 
\newcommand*{\infnro} {INFN-Roma La Sapienza, Piazzale A. Moro 2, I-00185 Roma, Italy} 
\newcommand*{\oab} {INAF-Osservatorio Astronomico di Brera, Via E. Bianchi 46, I-23807 Merate(LC), Italy} 

\author{P.W. Cattaneo}
\affiliation{\infnpv} 
\author{A. Rappoldi}
\affiliation{\infnpv} 
\author{A. Argan}
\affiliation{\inafro} 
\author{G. Barbiellini}
\affiliation{\infnts} 
\author{F. Boffelli}
\affiliation{\infnpv} 
\author{A. Bulgarelli}
\affiliation{\inafbo} 
\author{B. Buonomo}
\affiliation{\infnlnf} 
\author{M. Cardillo}
\affiliation{\inafro} 
\author{A.W. Chen}
\affiliation{\uniwits} 
\author{V. Cocco}
\affiliation{\inafro} 
\author{S. Colafrancesco}
\affiliation{\uniwits} 
\affiliation{\oar} 
\author{F. D'Ammando}
\affiliation{\inafirabo} 
\author{I. Donnarumma}
\affiliation{\inafro} 
\affiliation{\asi} 
\author{A. Ferrari}
\affiliation{\cifs} 
\author{V. Fioretti}
\affiliation{\inafbo} 
\author{L. Foggetta}
\affiliation{\infnlnf} 
\author{T. Froysland}
\affiliation{\unirotv} 
\affiliation{\cifs} 
\author{F. Fuschino}
\affiliation{\inafbo} 
\author{M. Galli}
\affiliation{\eneabo} 
\author{F. Gianotti}
\affiliation{\inafbo} 
\author{A. Giuliani}
\affiliation{\inafmi} 
\author{F. Longo}
\affiliation{\units} 
\affiliation{\infnts} 
\author{F. Lucarelli}
\affiliation{\ssdc} 
\affiliation{\oar} 
\author{M. Marisaldi}
\affiliation{\bergen} 
\affiliation{\inafbo} 
\author{G. Mazzitelli}
\affiliation{\infnlnf} 
\author{A. Morselli}
\affiliation{\infnrotv} 
\author{F. Paoletti}
\affiliation{\ewrsd} 
\affiliation{\inafro} 
\author{N. Parmigiani}
\affiliation{\inafbo} 
\author{A. Pellizzoni}
\affiliation{\oac} 
\author{G. Piano}
\affiliation{\inafro} 
\affiliation{\cifs} 
\author{M. Pilia}
\affiliation{\oac} 
\author{C. Pittori}
\affiliation{\ssdc} 
\affiliation{\oar} 
\author{M. Prest}
\affiliation{\unico} 
\author{G. Pucella}
\affiliation{\eneafr} 
\author{L. Quintieri}
\affiliation{\infnlnf} 
\author{S. Sabatini}
\affiliation{\inafro} 
\author{M. Tavani}
\affiliation{\inafro} 
\affiliation{\unirotv} 
\author{M. Trifoglio}
\affiliation{\inafbo} 
\author{A. Trois}
\affiliation{\oac} 
\author{P. Valente}
\affiliation{\infnro} 
\author{E. Vallazza}
\affiliation{\infnts} 
\author{S. Vercellone}
\affiliation{\oab} 
\author{F. Verrecchia}
\affiliation{\ssdc} 
\affiliation{\oar} 
\author{A. Zambra}
\affiliation{\inafmi}

\date{\today}

%\abstract
\begin{abstract}
{
AGILE is a mission of the Italian Space Agency (ASI) Scientific Program
dedicated to $\gamma$-ray astrophysics, operating in a low Earth orbit
since April 23, 2007.
It is designed to be a very light and compact instrument, capable of
simultaneously detecting and imaging photons 
in the \SIrange{18}{60}{\kilo\electronvolt} X-ray energy band 
%with a large field of view (FOV $\approx$\SI{1}{\steradian})
and in the \SI{30}{\MeV}--\SI{50}{\GeV} $\gamma$-ray energy 
%(FOV \SI{\approx 2.5}{\steradian})
with a good angular resolution ($\approx \ang{1} @$ \SI{1}{\GeV}).
The core of the instrument is the Silicon Tracker 
complemented with a CsI calorimeter and a AntiCoincidence system forming
the Gamma Ray Imaging Detector (GRID).
}
{Before launch, the GRID needed on-ground calibration with a tagged $\gamma$-ray beam 
to estimate its performance and validate the Monte Carlo simulation.}
{The GRID was calibrated using a tagged $\gamma$-ray beam with energy up to \SI{500}{\MeV}
at the Beam Test Facilities at the INFN Laboratori Nazionali di Frascati.
}
{These data are used to validate a GEANT3 based simulation by 
comparing the data and the Monte Carlo simulation by measuring the angular 
and energy resolutions.}
{The GRID angular and energy resolutions obtained using the beam agree well with the
Monte Carlo simulation. 
Therefore the simulation can be used
to simulate the same performance on-flight with high reliability.}

\end{abstract}

\correspondingauthor{P.W. Cattaneo}
\email{paolo.cattaneo@pv.infn.it}

%\maketitle

\section{Introduction}
\label{sec:intro}
AGILE (Astro-rivelatore Gamma a Immagini LEggero) is a Small Scientific
Mission of the Italian Space Agency (ASI) dedicated to high-energy 
$\gamma$-ray astrophysics \citep{agimis,agimis2}, composed of a Gamma Ray
Imager Detector (GRID) sensitive in the energy range 
\SI{30}{\MeV}--\SI{50}{\GeV} \citep{sttb,st1}, 
a hard X-ray imager (Super-AGILE) sensitive in the energy range \SIrange{18}{60}{\kilo\electronvolt} \citep{superagile}
and a Mini-Calorimeter (MCAL) sensitive to $\gamma$-rays and charged particles in the energy range
\SI{350}{\keV} to \SI{100}{\MeV} \citep{minical}.
At the core of the GRID, there is a Silicon Tracker (ST) for detection of $\gamma$-rays 
through pair production.

A correct interpretation of the GRID measurements relies on a precise calibration of the instrument.
This is based on a combination of on-ground calibration, Monte Carlo (MC) simulation and on-flight 
calibration. 
The goal of the present paper is to validate the MC simulation of the GRID 
by comparing the data and the MC simulation of the on-ground calibration with a $\gamma$-ray
tagged beam with energy up to \SI{500}{\MeV}.
For in-flight calibration see \citealt{chenspie,chenaa}.

\section{The AGILE instrument}

\begin{figure}[htb]
\begin{center}
\includegraphics[width=0.70\textwidth]{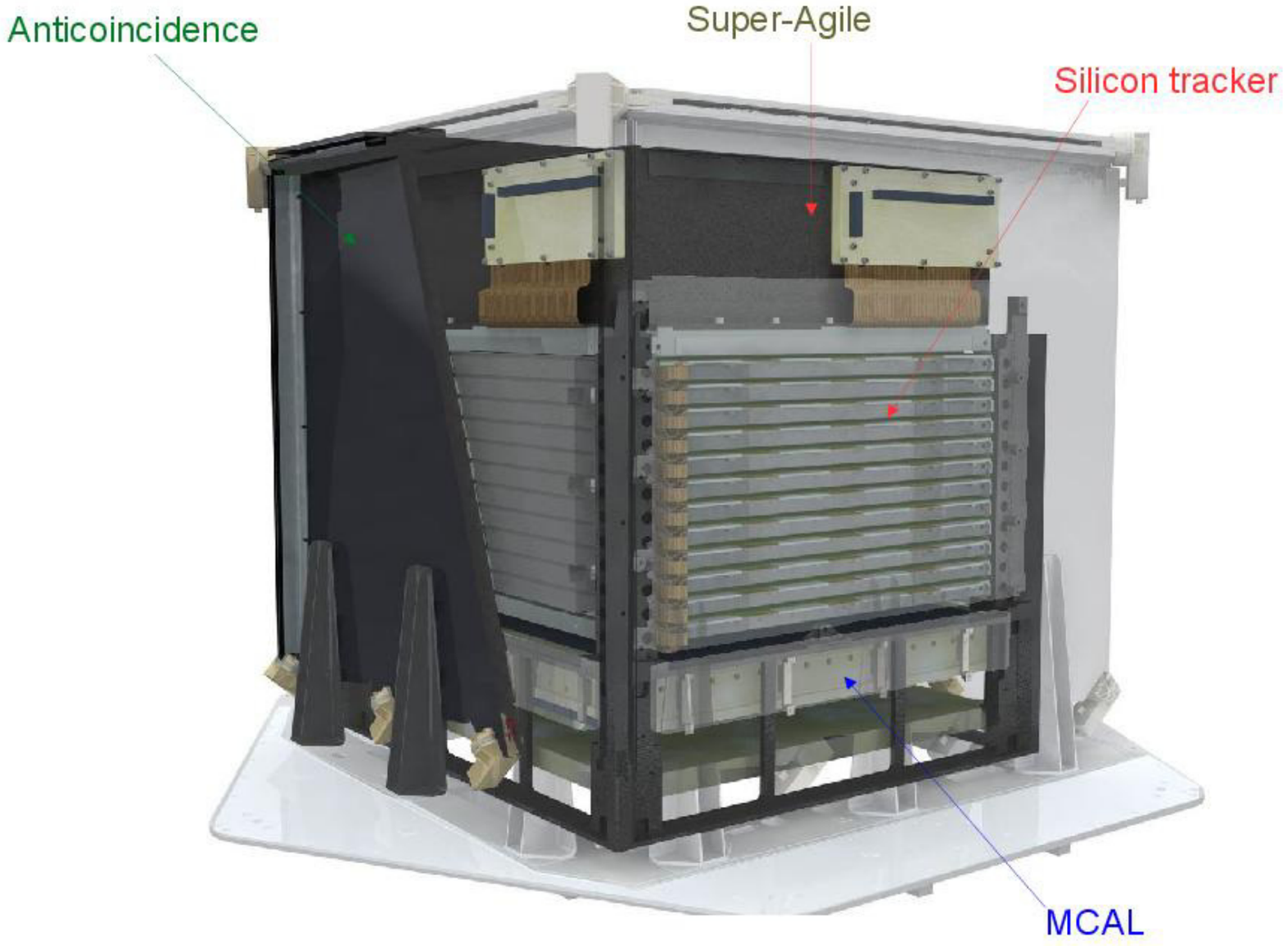}
\caption{Sketch of the AGILE payload, not in scale. The main components are shown: 
the plastic scintillator Anticoincidence, the Silicon-tungsten tracker, 
the CsI bars MiniCalorimeter, the hard X-ray imager Super-Agile based on a silicon 
detector plane and a tungsten mask.}
\label{agiledet}
\end{center}
\end{figure}

The AGILE scientific payload (sketched in its main components in Fig.~\ref{agiledet}) consists of 
three instruments with independent detection capability.

The Gamma-Ray Imaging Detector (GRID) consists of a Silicon-Tungsten converter-tracker (ST) with excellent spatial
resolution and good timing capability,
a shallow (\num{1.5} $\mathrm{X_0}$ on-axis) Caesium Iodide MCAL and an AntiCoincidence system
(AC) made of plastic slab \citep{ac}. It has an unprecedented large field of view (FOV) covering 
\SI{\approx 2.5}{\steradian}, almost 1/4 of
the entire sky, in the energy range \SI{30}{\MeV}--\SI{50}{\GeV}.

The hard X-ray imager (Super-AGILE) is a coded-masked system made of a silicon detector plane and
tungsten mask above it designed to image photons over a large FOV
($\approx$ \SI{1}{\steradian}).

MCAL operates also stand alone in ``burst mode" covering the range \SI{350}{\kilo\electronvolt} to
\SI{100}{\MeV} to detect GRB and other $\gamma$-ray transients.

%%%%%%%%%%%%%%%%%%%%%%%%%%%%%%%%%%%%%%%%%%%%%%%%%%%%%%%%%%%%%%%%%%%%%%%%%%%%%%%%%%%%%%%%

\subsection{The Silicon Tracker}

The core of the GRID is the ST with the task of converting the incoming $\gamma$-rays 
and measuring the trajectories of the resulting \eepair\ pair \citep{st1,sttb}.
The $\gamma$-rays convert in the W (Si) 
layers in \eepair\ pairs, which are subsequently detected by the silicon 
microstrip detectors.

The ST consists of 12 trays with distance between middle-planes equal to \SI{1.9}{\cm}.
The first 10 trays consists of pairs of single sided Si microstrip planes with strips orthogonal to each other
to provide 3D points followed by a W converter layer \SI{245}{\um} thick
(corresponding to 0.01(Si)+0.07(W) $\mathrm{X_0}$). The last two trays have no W converter layer since the ST trigger requires
at least three Si planes to be activated.

The base detector unit is a tile of area \SI[product-units=single]{9.5 x 9.5}{\cm\squared}, thickness 
\SI{410}{\um} and strip pitch \SI{121}{\um}. Four tiles are bonded together to 
create a `ladder' with \SI{38.0}{\cm} long strip.
Each plane of the ST consists of four ladders. The readout pitch is \SI{242}{\um}.

The ST analogue read-out measures the energy deposited on every second strip; 
that is `readout' and  `floating' strips alternate. This configuration enables to obtain a good
compromise between power consumption and position resolution, the latter being 
$\sim$\SI{45}{\um} for perpendicular tracks \citep{sttb}. 

Each ladder is readout by three TAA1 ASICs, each operating 128 channels at
low noise, low power configuration (\SI{<400}{\micro\watt}/channel), self-triggering capability and analogue readout.
The total number of readout channels is 36864.

\subsection{The GRID simulation}

The GRID is simulated using the GEANT 3.21 package \citep{geant}.
This package provides for a 
detailed simulation of the materials and describes with high precision the passage 
of particles through matter including the production of secondary particles. It 
provides the user with the possibility of describing the response of the active 
sections of the detector.
The GRID detector simulation used in the test beam configuration is used also
to simulate the configuration with the detector mounted on the spacecraft.

In a microstrip detector with floating strips a critical aspect of the simulation 
is the sharing of the charge collected on the strips to the readout channels \citep{cc1990}. 
These sharing coefficients are estimated by using test beam data 
to reproduce the observations \citep{sttb}.

The Cartesian coordinate system employed in the following has the origin in the centre 
of the AC plane on the top, the x and y axis parallel to the orthogonal strips of the ST
and the z axis pointing in the direction of MCAL. In the following the corresponding spherical
coordinate system defined by the polar angle $\Theta$ and azimuthal angle $\Phi$ will be extensively used.

\subsection{Reconstruction filter}

The reconstruction filter processes the reconstructed hits in the ST layers with 
the goal of providing a full description of the events.
The reconstruction filter has several, sometimes conflicting, goals:
reconstructing tracks, estimating their directions and energies,
combining them to identify $\gamma$-ray,
rejecting background
hits from noise and charged particle escaping the AC veto 
and providing an estimation of the probability that 
the measured tracks originate by a pair converted $\gamma$-ray.
The filter used in this analysis, FM3.119 \citep{bulgarelli2010,chen2011}, is the result of an optimisation
between those requirements
and has been used in all AGILE scientific analyses.

\subsubsection{Direction reconstruction}

The reconstruction of the $\gamma$-ray direction, defined 
by the polar angle $\Theta$ and azimuthal angle $\Phi$ 
with respect to the AGILE coordinate system,
is based on the process of pair
production and is obtained from the identification and the analysis of the
\eepair\ tracks stemming from a common vertex. 
At each tray the microstrips on the silicon layers measure separately 
the coordinates x and y of the hits.
The first step of the event analysis requires to find the two tracks among the
possible associations of the hits detected by the ST layers.

The second step consists in a linear fit of the hits associated to each track.
This task is complicated by the electrons moving along not straight line trajectory 
because of the multiple scattering.
These steps are performed separately for the hits corresponding to the x and y
coordinates producing four tracks, two for each projection. 
The direction in three dimension is obtained associating
correctly the two projections of each track \citep{arem}.

In order to fit the tracks, a Kalman filter smooth algorithm \citep{kalman,
frukal} has been developed \citep{giukal}. This technique progressively updates the
track candidate information during the track finding process, predicting as
precisely as possible the next hit to be found along a trajectory. This
capability is used to merge into an unique recursive algorithm the
track-finding procedure and the fitting of the tracks parameters.
It is therefore possible to associate at each fitted track a total $\chi^2$ given
by the sum of the $\chi^2$ of each plane of the ST. If the combination of the ST hits
gives rise to more than two possible tracks, the $\chi^2$ is used to
choose the best pair of reconstructed tracks corresponding to the
\eepair\ pair originating from the $\gamma$-ray conversion vertex.

\subsubsection{Energy Measurement}
\label{enermeas}

The track reconstruction algorithm is based on a special implementation 
of the Kalman filter that estimates the energy of the single tracks 
on the basis of the measurement of the multiple scattering
between adjacent planes.

A quantitative estimation of the single track energy resolution can be obtained 
considering the Gaussian approximation of the standard deviation of the distribution of the 
3D multiple scattering angle ($p$ is the particle momentum, $\beta$ the speed,
$t/\mathrm{X_0}$ the thickness of the material expressed in term of radiation length) and the
ultrarelativistic approximation ($\beta=c$ and $pc=E$)

\begin{equation}
\sigma_\mathrm{MS}(\theta) = \frac{\SI{19.2}{\MeV}}{c\beta p} \sqrt{\frac{t}{\mathrm{X_0}}} \sim 
\frac{\SI{19.2}{\MeV}}{cE} \sqrt{\frac{t}{\mathrm{X_0}}}.
\label{mstheta}
\end{equation}
Therefore to estimate the error on energy measurement we can write 

\begin{eqnarray}
\Delta(\sigma_\mathrm{MS}) &=& \frac{\Delta(E)}{E} \frac{\SI{19.2}{\MeV}}{cE} 
\sqrt{\frac{t}{\mathrm{X_0}}} \nonumber \\
\frac{\Delta(E)}{E} &=& \frac{\Delta(\sigma_\mathrm{MS})}{\sigma_\mathrm{MS}} 
\label{releeneerr}
\end{eqnarray}

The trajectory of a charged particle subject to multiple scattering crossing $N$ measurement planes 
is a broken line consisting of $N-1$ segments that define $N_\mathrm{MS} = N - 2$ scattering angles. 
The particle direction projected on one coordinate between the planes $k$ and $k+1$ at a distance $L$ is given by
$(x_{k+1} - x_k)/L$; this variable has an associated measurement error due to the position error $\sigma(x_k)$.
If the measurement of the particle direction between measurement planes is dominated
by the change of direction due to multiple scattering (and not by the position measurement error),
it provides $N_\mathrm{MS}$ sampling of the multiple scattering distribution,
assuming that the particle energy is constant throughout the tracking.
The error associated to the measurement of the standard deviation of the Gaussian distribution 
in Eq.\ref{mstheta} is
\[
\Delta (\sigma_\mathrm{MS}) = \frac{1}{\sqrt{2N_\mathrm{MS}}} \sigma_\mathrm{MS}
\]
From Eq.\ref{releeneerr} we obtain
\begin{equation}
\Delta(\log E) = \frac{\Delta(E)}{E} = \frac{\Delta(\sigma_\mathrm{MS})}{\sigma_\mathrm{MS}} = \frac{1}{\sqrt{2N_\mathrm{MS}}},
\label{eeneerr}
\end{equation}
hence the relative error is independent from the energy and from the thickness of
scattering material and therefore from the angle relative to the planes and depends only 
on the number of hits.
In average, having 12 planes, the maximum value for $N_\mathrm{MS}$ is 10. The average value is half this 
value $\sim 5$. Therefore we expect $\Delta(\log E) \sim {1}/\sqrt{2\times 5} \sim 0.316$.

The high-energy limit of this approach is reached when the size of the deviation 
from the extrapolated trajectory due to multiple scattering is comparable to the 
error in position measurement in the measurement plane. From \citealt{sttb} the measurement 
error of the single coordinate is $\sigma_\mathrm{x} \SI{\sim 45}{\um}$ and, therefore,
on the 2D distance in the tracking plane $\sigma_\mathrm{xy} \sqrt{2}\times\SI{\sim 45}{\um} 
\SI{\sim 64}{\um}$, while L=\SI{1.9}{\cm}.
Hence the limit on the capability of measuring the direction of a track segment, 
which has two vertices, is $\sigma_\mathrm{xy}/L \sim 0.0064 \sqrt{2}/1.9 \sim$\SI{4.8e-3}{\radian}. 
Recalling that the thickness of a tray is \SI{0.08}{X_0} and that the angular deviation due to
multiple scattering is measured from the different between the directions of the track segments, 
from Eq.\ref{mstheta} the particle
energy for which the multiple scattering angle is equal to the direction error due to 
the finite position resolution of the planes is \SI{\sim 0.8}{\GeV}. 
Therefore in the energy range available in this test the assumption that the multiple 
scattering contribution dominates over measurement errors is always true.

\subsubsection{Event classification }

On the basis of the reconstruction results, 
each event is classified by the filter as a likely $\gamma$-ray (G), limbo (L),
a particle (P) or a single-track event (S). 
Limbo events look like G events but not exactly, for example there are additional hits.
In practice, all scientific
analyses other than pulsar timing and $\gamma$-ray bursts have
used only G events. Therefore we concentrate on G events
in the following.

\section{The Gamma-Ray Calibrations}

\subsection{Calibration goals}
\label{merit}

The goal of the calibration is to estimate the instrument response
functions by means of exposure to controlled $\gamma$-ray beams. 
The ideal calibration beam provides a flux of $\gamma$-rays, monochromatic in 
energy and arriving as a planar wave uniformly distributed on the instrument surface, 
with properties known to an accuracy better than the resolving power of the instrument.

The figures of merit to be evaluated and compared are 
the Energy Dispersion Probability (EDP),
the Point Spread Function (PSF) and the effective Area (\aeff): all of them 
depend on the incoming $\gamma$-ray energy $\egamma$ and direction $(\Theta,\Phi)$. 

The EDP is the energy response of the GRID detector to a monochromatic planar wave
with energy $\egamma$ and direction $(\Theta,\Phi)$.
It is a function of $\egamma$ and of $(\Theta,\Phi)$, defined as
\begin{equation}
EDP(\egamma,\Theta,\Phi) = \frac{1}{N(\egamma,\Theta,\Phi)}
  \frac{dN(\egamma,\Theta,\Phi)}{d\egrid}.
\label{edp}
\end{equation}
where $\egrid$ is the energy measured by the GRID.

The PSF is the response in the angular domain of the GRID detector to a planar wave. 
If $\gamma$-rays impinge uniformly on the GRID with fixed direction $(\Theta,\Phi)$,
the detector measures for each $\gamma$-ray a direction $(\Theta^\prime,\Phi^\prime)$.
Assuming this function to be azimuthally symmetric, 
and defining the three-dimensional angular distance $\theta$ between the true 
and measured directions\footnote{$\cos(\theta)$ is the scalar product of the versors identified 
by $(\Theta,\Phi)$ and $(\Theta^\prime,\Phi^\prime)$}, it is defined as
\begin{equation}
  PSF(\theta, \egamma,\Theta,\Phi)d\theta = 2\pi \sin(\theta) P(\theta,\egamma,\Theta,\Phi) d\theta
\end{equation}
where $P(\theta,\egamma,\Theta,\Phi)$ is the
probability distribution per steradian of measuring an incoming
$\gamma$-ray at a given angular distance $\theta$ from its true direction \citep{sabatini2010}.

The \aeff\ is a function of $\egamma$ and of $(\Theta,\Phi)$, it is defined as 
\begin{equation}
  \aeffm (\egamma, \Theta, \Phi) = \int_A \epsilon(\egamma, \Theta, \Phi, \overline{x}) da 
\end{equation}
where the integral over the two-dimensional coordinate variables covers the detector area 
$A$ intercepted by the incoming $\gamma$-ray direction
and $\epsilon(\egamma, \Theta, \Phi, \overline{x})$ is the detection efficiency dependent 
on the energy and direction of the incoming $\gamma$-ray and on the position on the detector.

The GRID was calibrated at the INFN Laboratory of Frascati (LNF) in
the period 2-20 November 2005, thanks to a scientific collaboration between AGILE
Team and INFN-LNF.

\subsection{Calibration strategy}

The goal of the GRID calibration is to reproduce with adequate statistic 
in the controlled environment of the laboratory the $\gamma$-ray interactions 
under space conditions. The total number of
required incident $\gamma$-rays $\mathrm{N_T}$, not necessarily interacting
within the GRID, depends on the expected counting statistics for bright astrophysical
sources acquired for typical exposures.
With the goal of achieving statistical errors due to the calibration negligible
compared to those in flight, we require a number of events about four times larger
than the brightest source.

%It is important to note that the determination of the GRID \aeff 
%requires the production and tagging of all incident $\gamma$-rays, independently
%whether they interact or not within the GRID.
%
The canonical source of in-flight calibration is the Crab Nebula with an
integrated average total $\gamma$-ray flux (nebula and pulsar) of 
$I_\mathrm{c}(\egamma\SI{> 100}{\MeV})$ = \SI{220e-8}{ph\per\cm\squared\per\second} 
\citep{catalog1} and the calibration statistics is 
estimated based on its flux and the AGILE detection capability \citep{agimis2}.
The Crab Nebula integral $\gamma$-ray intensity flux as a function of the $\gamma$-ray 
energy $\egamma$ expressed in \si{\MeV} is
$I_\mathrm{c}(\egamma\SI{> 100}{\MeV})$ = $\egamma^{-1.015}$ \SI{2.36e-4}{ph\per\cm\squared\per\second}.

For a given incident direction $(\Theta$,$\Phi)$, the number of required incident $\gamma$-rays 
with $\egamma>E_\mathrm{th}$ 
is $N(E_\mathrm{th},\Theta,\Phi) = \alpha t_\mathrm{exp} \eta A_\mathrm{geom} I_\mathrm{c}(\egamma)$, 
where $\alpha=4$ is the factor required to reduce to negligible levels the statistical error 
due to calibration, $\eta$ is the Earth occultation efficiency
(typically $\eta = 0.45$), $t_\mathrm{exp}$ the exposure time (typically $t_\mathrm{exp} = 2$ weeks) and
$A_\mathrm{geom}$ the geometric area of the GRID ($A_\mathrm{geom} = \SI{1600}{\cm\squared}$).

For typical values and the GRID geometry, we obtain the integrated required
number of incident $\gamma$-ray for $\egamma > \SI{30}{\MeV}$ for a given combination of
$(\Theta,\Phi)$:
\begin{equation}
%N(\SI{30}{\MeV}, \Theta,\Phi) = \SI{1.8e-4}{ph}.
N(\SI{30}{\MeV}, \Theta,\Phi) = \SI{2.6e4}{ph}.
\label{neg30}
\end{equation}

\subsection{Calibration set up}

The tagged $\gamma$-ray beam used for the calibration 
has been already described in detail in \citealt{btfagile}.
In the following the most relevant elements are briefly summarised.

\subsubsection{The Beam Test Facility}

\begin{figure}[htb]
\begin{center}
\includegraphics[width=0.48\textwidth,angle=0]{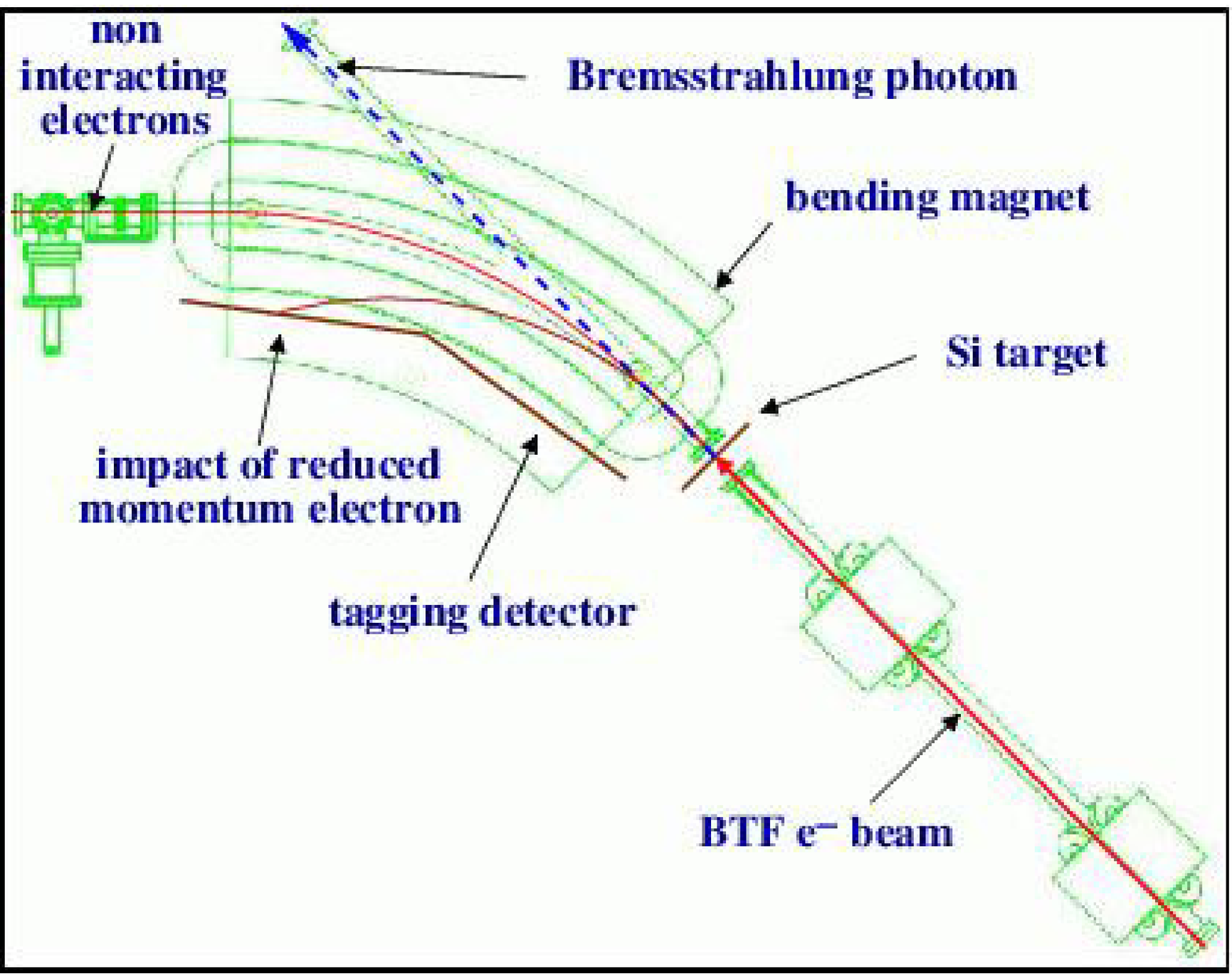}
\caption{Schematic view of the $\gamma$-ray line: the target, the spectrometer magnet and the 
Photon Tagging System (PTS).}
\label{lineafascio}
\end{center}
\end{figure}

For the GRID calibration we used the Beam Test Facility (BTF) 
in the Frascati DA$\Phi$NE collider complex, which includes a LINAC at 
high electron and positron currents, an accumulator
of \eepair\ and two accumulation rings at \SI{510}{\MeV}.
The e$^+$/e$^-$ beam from the LINAC is led into the accumulation ring to be
subsequently extract and injected in the Principal Ring. When the system
injector does not transfer the beams to the accumulator, the beam from LINAC can
be transported in the test beam area through a dedicated transfer line: 
the BTF line (Fig.~\ref{lineafascio}).
The BTF can provide a collimated beam of electrons or positrons in the energy range
\SIrange{20}{800}{\MeV} with a pulse rate of \SI{49}{\hertz}. 
The pulse duration can vary from \SIrange{1}{10}{\nano\second} and the average 
number of particles for bunch from 1 to $10^{10}$.
In Fig.~\ref{lineafascio} a schematic view of the calibration set up is shown.

%%%%%%%%%%%%%%%%%%%%%%%%%%%%%%%%%%%%%%%%%%%%%%%%%%%%%%%%%%%%%%%%%%%%%%%%%%%%%%%%%%%%%%%%%%
%%%%%%%%%%%%%%%%%%%%%%%%%%%%%%%%%%%%%%%%%%%%%%%%%%%%%%%%%%%%%%%%%%%%%%%%%%%%%%%%%%%%%%%%%%%%

The BTF can be operated in two ways
\begin{itemize}
\item a LINAC mode operating when DA$\Phi$NE is off with a tunable energy in the 
range \SIrange{50}{750}{\MeV} and an efficiency around 0.9
\item a DA$\Phi$NE mode operating when DA$\Phi$NE is on with a fixed energy of 
\SI{510}{\MeV} and an efficiency around 0.6
\end{itemize}

\subsubsection{Target }

Beam electrons cross perpendicularly the thin silicon microstrip detectors 
acting as target, producing $\gamma$-rays in the energy 
range of interest to the GRID by Bremsstrahlung; 
subsequently a dipole magnet bends away electrons while $\gamma$-rays 
follow straight trajectories and can reach the GRID instrument.

The target consists of two pairs of single-sided silicon microstrip detectors 
\SI{0.41}{\mm} thick of area \SI[product-units=single]{8.75 x 8.75}{\cm\squared}. 
Each detector includes 
384 strips with \SI{228}{\um} pitch. The target has two functions: measuring the passage 
of each electron and causing the emission of Bremsstrahlung $\gamma$-rays.

\subsubsection{Photon Tagging System (PTS)}

Our team developed and installed in the BTF area a Photon Tagging System (PTS)
for the detection of the electrons that produce Bremsstrahlung $\gamma$-rays 
in the target \citep{btfagile}.
In absence of significant energy loss due to Bremsstrahlung, the electrons 
follow a circular trajectory inside a guide inserted in the dipole magnet
without interacting, as sketched in Fig.~\ref{lineafascio}.
The internal wall of the guide is covered by silicon microstrip detectors;
therefore, in presence of emission of Bremsstrahlung $\gamma$-rays of 
sufficient energy, the electrons follow a more curved trajectory and hit 
one of the microstrip detectors in a position correlated to the energy loss. 
After calibrating with MC simulation the relation between PTS hit position
and $\gamma$-ray energy, the PTS position measurement provides an estimator of the
$\gamma$-ray energy $\epts$.

\subsubsection{Trade-off on the number of e$^-$/bunch}

The overall GRID performance should be tested in a single-photon regime without 
simultaneous multi-photon interactions, which are not represent astrophysical conditions 
and cannot be easily identified during the calibration.
Multi-photon interactions represent an intrinsic noise that may bias significantly the measure
of the \aeff\ and of the EDP.

Ideally, one $\gamma$-ray should be emitted by one electron crossing the target, but both
the actual number of electrons in each bunch and the number of emitted Bremsstrahlung $\gamma$-rays
are stochastic variables. Therefore multi-photon emission cannot be eliminated completely.

For the calibration a compromise between BTF efficiency and calibration accuracy
had to be found. In the DA$\Phi$NE mode with 5 e$^-$/bunch the fraction
of events with multiple $\gamma$-rays having $\egamma$ \SI{>20}{\MeV} 
is $\sim 5$\%. This uncertainty 
is comparable to the accuracy requirement on the measurement of \aeff.
On the other hand, the DA$\Phi$NE mode with 1
e$^-$/bunch is consistent with an accuracy requirements $\sim 1$\%\ but reduces the
number of configurations/directions that can be calibrated due to the constraint of 
available time. 

Taking into account the above considerations, the 
configuration with 3 e$^-$/bunch was chosen as compromise between 
reduced multi-photon events and sufficient statistics.

\subsubsection{Mechanical Ground Support Equipment}

Our team developed and installed in the BTF area a Mechanical Ground Support
Equipment (MGSE) that was used for the AGILE calibration \citep{gianotti}.
The MGSE permitted the rotation and translation of AGILE relatively to the beam.
A rotation is parametrised by the polar angle $\Theta$ and azimuthal angle $\Phi$ 
of the beam direction with respect to the AGILE coordinate system.
Using the translation capability of the MGSE, for each run the beam was focused
in a predefined position on the GRID. 

\subsubsection{Simulation}

The overall system, including the beam terminal section, the
target, the bending magnet, the PTS and the GRID is simulated
in detail using GEANT 3.21 package \citep{coccomc2001, coccomc2002}.
This allows a direct comparison between the resolutions measured
in simulated and real data providing a validation of the MC simulations.
An improvement of the comparison between data
and MC is obtained by overlapping a uniform flux of low energy 
photons to the Bremsstrahlung $\gamma$-ray. These photons 
are beam related and are a background that cannot be precisely 
estimated and is tuned to match the experimental data.

\section{Data samples}

During the calibration campaign $\num{\sim 2.7e6}$ PTS events were detected, 
of which $\sim 40\%$ have interacted with the GRID; they impinged 
in different positions on the GRID for a set of predefined incident directions.
Polar angles $\Theta$ between \ang{0} and \ang{65} and azimuthal angles $\Phi$ between 
\ang{0} and \ang{315} were tested. Table~\ref{data-samples} presents a detailed account 
of the data samples versus the beam incident angles. 
For each direction, the number of PTS events is significantly larger, and also the number 
of tagged $\gamma$-rays is at least comparable, than the number in 
Eq.~\ref{neg30} guaranteeing small statistical errors.

\begin{figure}[htb]
\begin{center}
\mbox{\begin{tabular}[t]{cc}
\subfigure[]{\includegraphics[width=0.23\textwidth]{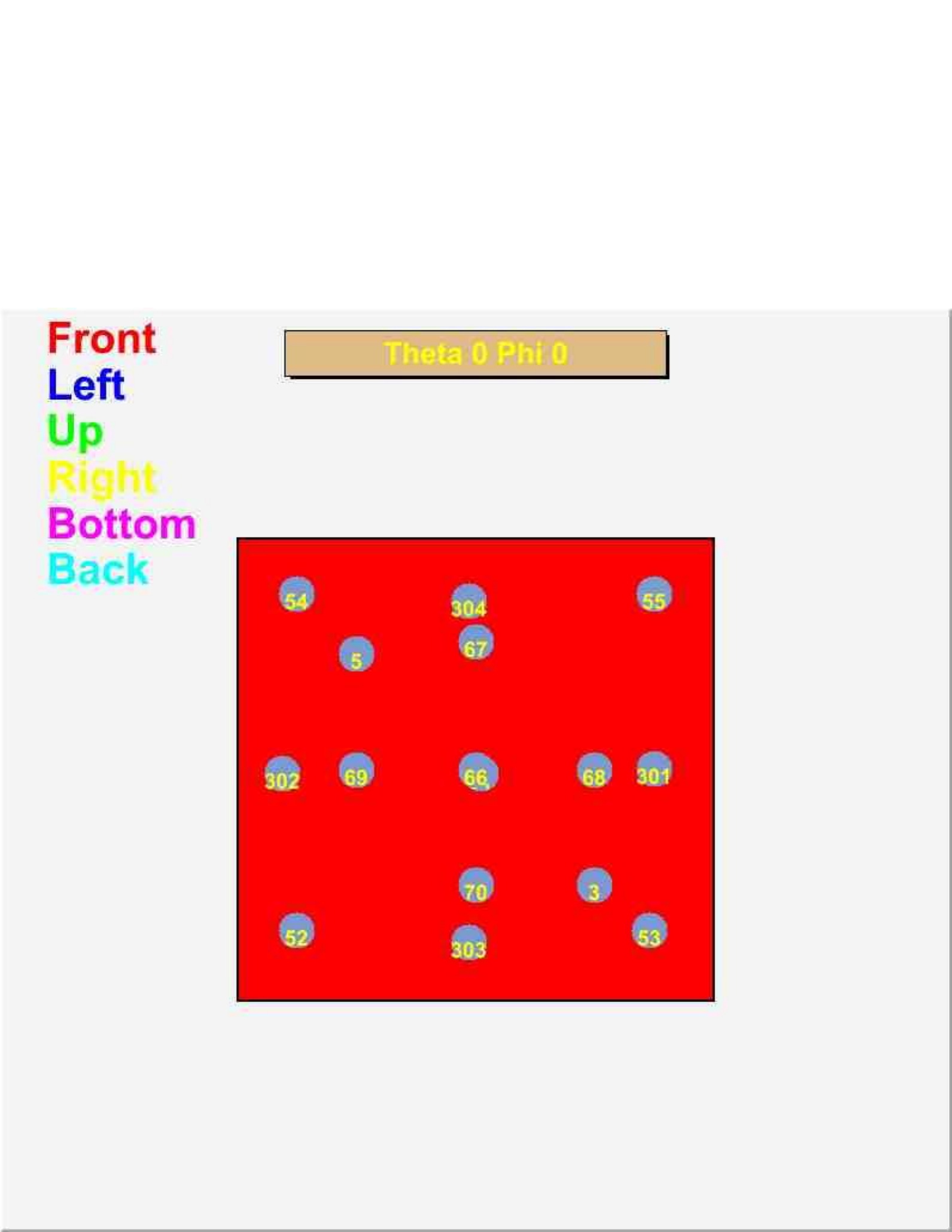}
} &
\subfigure[]{\includegraphics[width=0.23\textwidth]{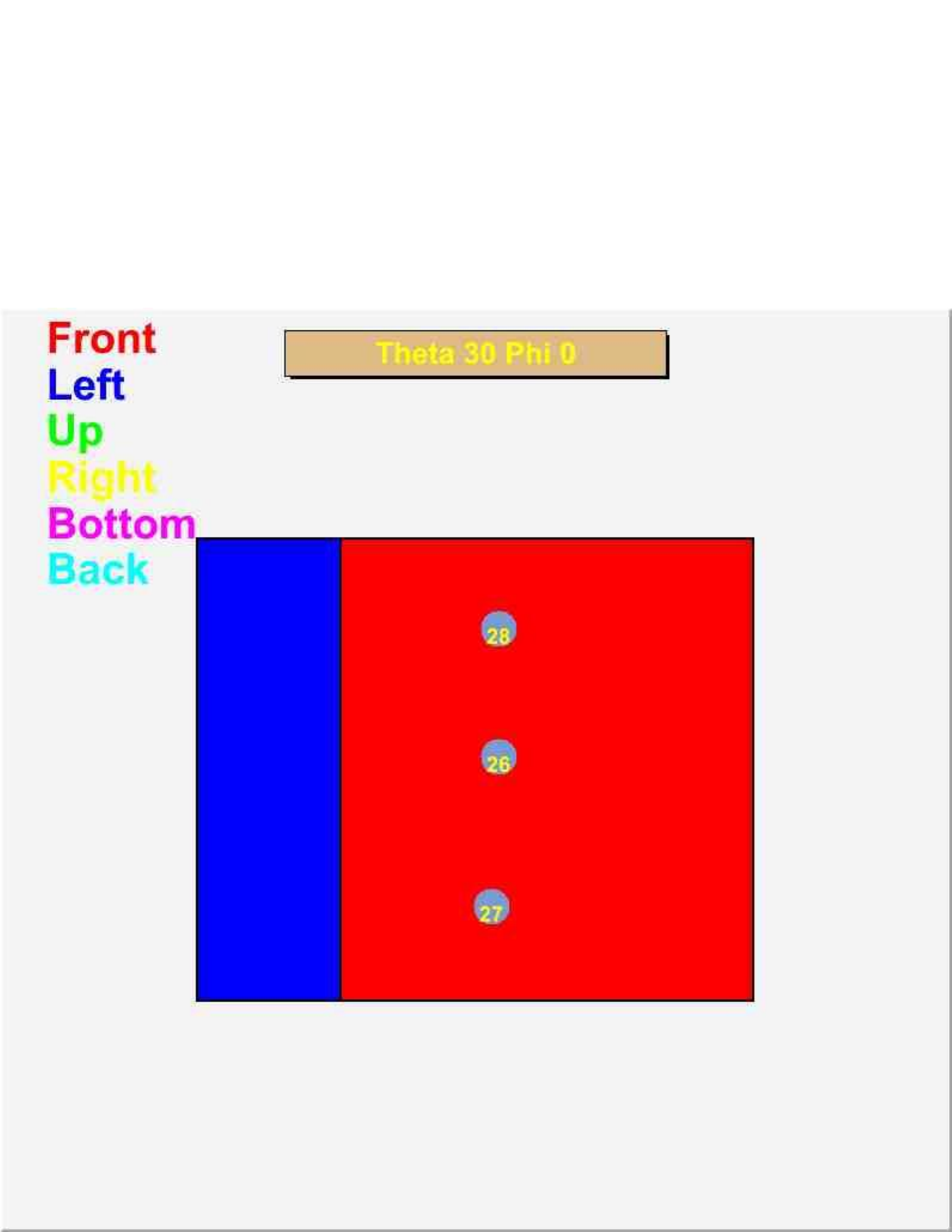}
} \\
\subfigure[]{\includegraphics[width=0.23\textwidth]{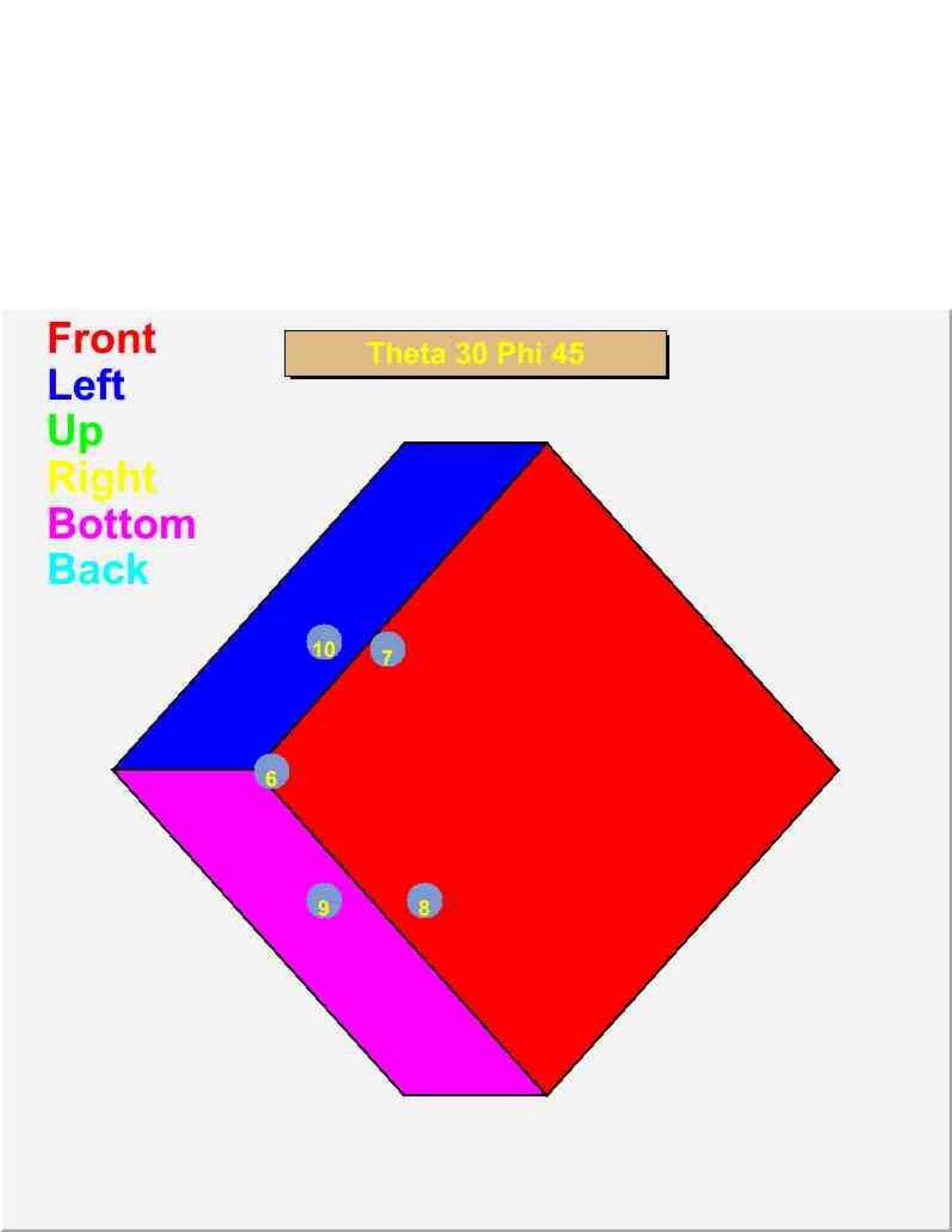}
} &
\subfigure[]{\includegraphics[width=0.23\textwidth]{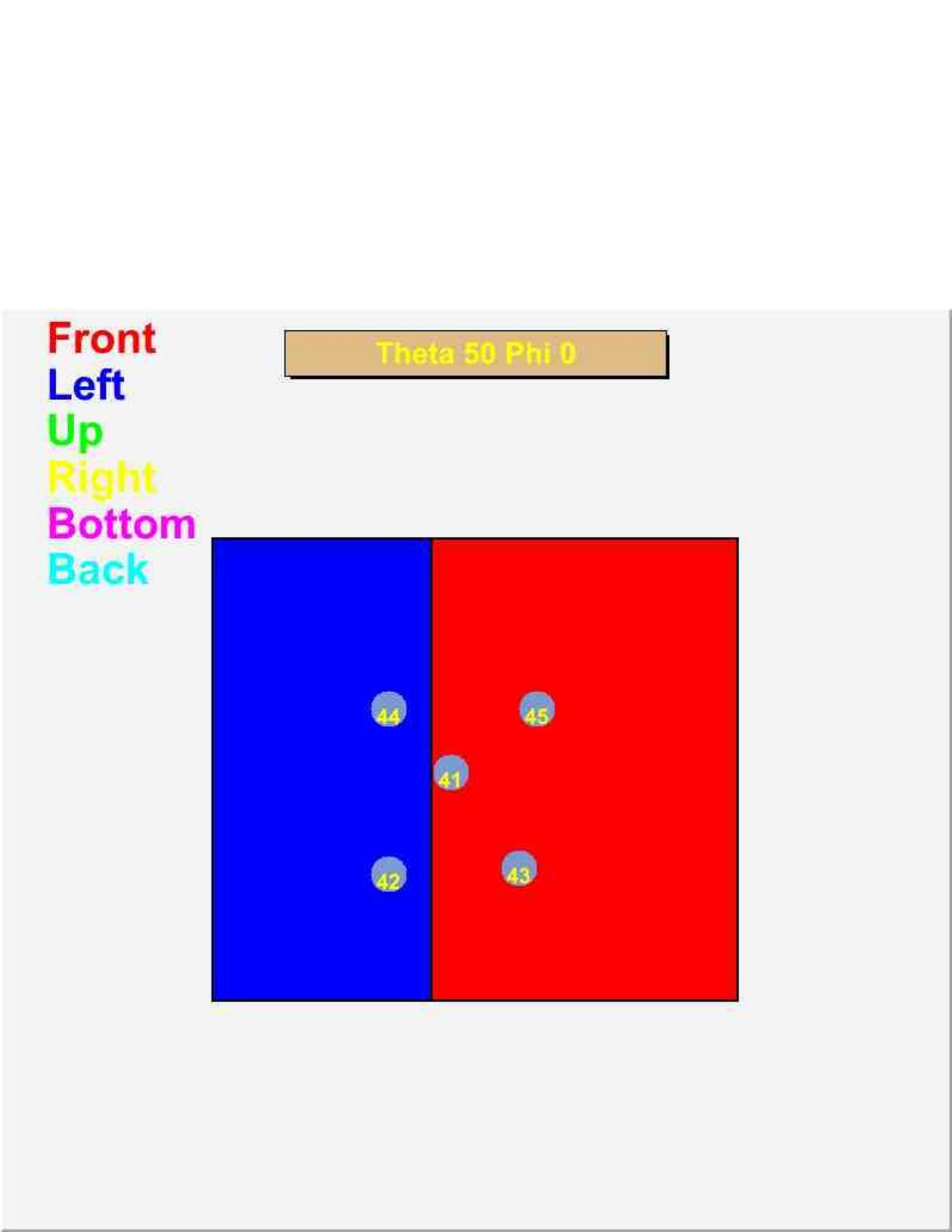}
} \\
\end{tabular}}
\caption{Beam incident positions for a) $\Theta$\ang{=0}, $\Phi$\ang{=0}
b) $\Theta$\ang{=30}, $\Phi$\ang{=0}, c) $\Theta$\ang{=30}, $\Phi$\ang{=45}, d) $\Theta$\ang{=50}, $\Phi$\ang{=0}.}
\label{spill}
\end{center}
\end{figure}

For each direction, a set of incident positions was selected to provide an approximate
coverage of the detector area; some examples are shown in Fig.~\ref{spill}.
It is nevertheless evident that the coverage is very approximate with the partial 
exception of the configuration $\Theta$=\ang{0}, $\Phi$=\ang{0}. 
This makes impossible a direct estimation 
of the \aeff\ that requires a uniform illumination of the instrument.

\begin{table}[!t]
\caption{Size of data samples taken in various configurations: in the first two columns
the polar and azimuthal angles of the AGILE axis with respect to the beam, 
in the third the number of events with a GRID trigger, in the fourth the number of events
with a PTS trigger, in the fifth the number of events with the GRID-PTS triggers in time,
in the sixth the ratio between the fifth and the fourth columns, representing, 
approximately, the GRID efficiency.}
\label{data-samples}
\centering
\begin{tabular}{|c|c|c|c|c|c|}
\hline
$\Theta$ & $\Phi$ & GRID events & PTS events & Tagged $\gamma$ & $\epsilon$(\%) \\
\hline
\hline
\ang{0}  & \ang{0} & 584873  & 682269 & 284726 & 41.7\\
\hline
\hline
\ang{30} & \ang{0} & 107132 & 133322 & 59751 & 44.8 \\
\hline
\ang{30} & \ang{45} & 332604 & 458633 & 192872 & 42.1 \\
\hline
\ang{30} & \ang{135} & 290792 & 379584 & 157908 & 41.6 \\
\hline
\ang{30} & \ang{225} & 154379 & 176824 & 70898 & 40.1 \\
\hline
\ang{30} & \ang{270} & 155636 & 180578 & 84477 & 46.8 \\
\hline
\ang{30} & \ang{315} & 63367 & 75048 & 33811 & 45.0 \\
\hline
\hline
\ang{30} & all & 1103910 & 1403989 & 599797 & 42.7 \\
\hline
\hline
\ang{50} &   \ang{0} & 219479 & 227402 & 108656 & 47.8 \\
\hline
\ang{50} & \ang{270} & 127389 & 135887 & 66647 & 49.0 \\
\hline
\ang{50} & \ang{315} & 164296 & 183198 & 85351 & 46.6 \\
\hline
\hline
\ang{50} & all & 511164 & 546487 & 260654 & 47.7 \\
\hline
\hline
\ang{65} &  \ang{0} & 49745 & 46369 & 22799 & 49.2 \\
%\hline
%\ang{65} & \ang{270}} & 41952 & 49564 & 1294 \\
\hline
\ang{65} & \ang{315} & 51359 & 40759 & 21509 & 52.8 \\
\hline
\hline
\ang{65} & all & 101104 & 87128 & 44308 & 50.8 \\
\hline
\hline
all & all & 2301051 & 2719873 & 1189485 & 43.7 \\
\hline
\end{tabular}
\end{table}

\section{Analysis of Events}

\subsection{The GRID Trigger}

A great challenge in $\gamma$-ray astronomy is due to the fact that cosmic
rays and the secondary particles produced by the interaction of cosmic rays in the
atmosphere produce in detectors a background signal much larger than that produced
by cosmic $\gamma$-rays. Therefore the development and the optimisation of the trigger
algorithms to reject this background have been of primary relevance for AGILE.
AGILE have a Data Handling system \citep{agile-DH-ieee-2008,agimis} which cuts a
large part of the background
rate through a trigger system, consisting of both hardware and software levels,
in order not to saturate the telemetry channel for scientific data.

The on-board GRID trigger is divided into three levels: two hardware (Level 1 and 1.5)
and one software (Level 2) \citep{giukal}. 
%The ST readout system consists in a synchronous part and an asynchronous 
%one to reduce the dead time of the instrument.
%The Level 1 trigger is given by the coincidence of 3 of 4 consecutive ST
%planes, to take into account the possibility of one-plain failure.
%According to the Level 1 trigger conditions, there must be at least three
%consecutive ST planes which give a signal in both views. With respect to
%trigger Level 1, we perform a further event selection by asking that accepted
%events show only one cluster of strips on the first hit ST plane in both views.
%This amounts to require that it be possible to identify a clean conversion
%vertex in both ST views. We also require no more than two clusters on the
%second hit ST plane, always in both views. In such a way we reject all
%events with spurious hits on the second plane. No further conditions are
%imposed on the third and subsequent planes.
The hardware levels accept only the events that produce some particular
configuration of fired anticoincidence panels and ST planes.
The main goal of the two hardware trigger levels is to select $\gamma$-rays
and provide a rejection factor of $\sim 100$ for background events. 
The software level provides an additional rejection of the background 
by a factor $\sim 5$.

The residual background events are
further reduced offline on ground by more complex software processing.
In the first step software algorithms are applied that analyse the
event morphology, based on cluster identification and topology to reduce the
background. 
A further processing step is necessary to eliminate the remaining background events, 
particularly those produced by albedo $\gamma$-rays which are discriminated by cosmic 
$\gamma$-rays only on the basis of their incoming direction.
(see \citealt{vercellone2008}
for a description of the AGILE on-ground data reduction).

\subsection{The PTS trigger }

The trigger for reading out the target and the PTS data was delivered by the accelerator 
and was independent from the PTS data themselves.
Therefore the PTS events were triggered independently from the GRID and 
written on a separate steam. Further selection based on the PTS data was performed 
offline by requiring that one or more strip clusters were identified. 
A cluster was defined as a set of neighbouring strips on the same detector 
above a predefined threshold. The signal over noise was high enough that the 
algorithm selected essentially all electrons crossing the detector \citep{btfagile}. 

\subsection{Event selection for data}

The PTS events and the GRID ones, which are written on two different streams,
are correlated offline exploiting the time tags recorded for both event streams.
The result is a subset of GRID events that can be associated, for each event, to the $\gamma$-ray 
energy estimated by the PTS $\epts$.
The fraction of events with a PTS trigger and without a GRID trigger provides
an estimation of the efficiency of the GRID.

\subsection{Event selection for MC}

MC sets are simulated reproducing precisely the beam parameters and the 
position and orientation of the GRID with respect to the beam as well as
online and offline trigger conditions. 
Event samples comparable in size to the data are generated.
The events are selected by requiring that the trigger conditions are
satisfied by PTS and GRID signals, 
so that direct comparison with data is possible.

\subsection{PTS as energy estimator}
\label{ptsenerest}

\begin{figure*}[htb]
\begin{center}
\mbox{\begin{tabular}[t]{cc}
\subfigure[]{\includegraphics[width=0.48\textwidth]{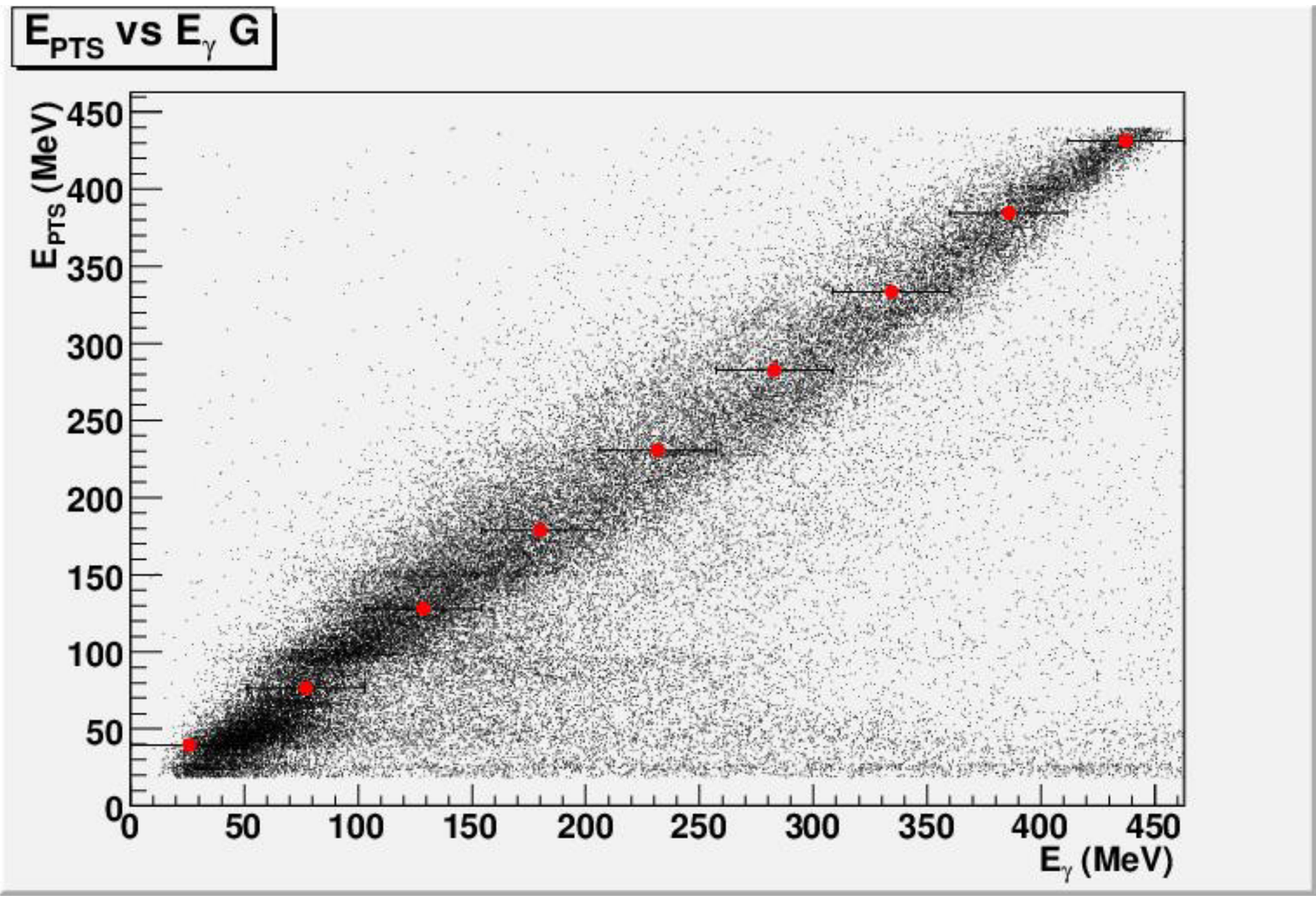} } &
\subfigure[]{\includegraphics[width=0.48\textwidth]{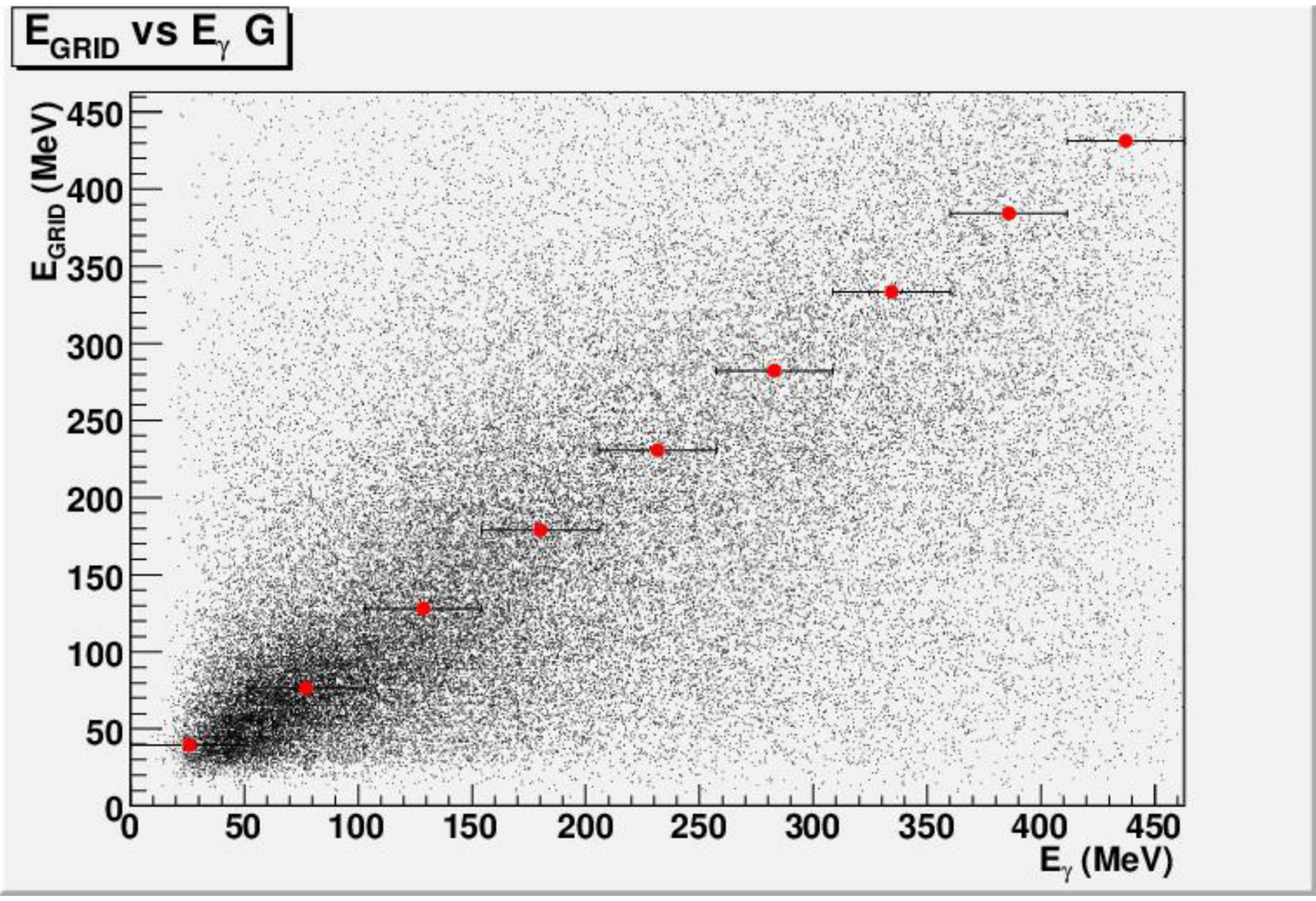} } \\
\end{tabular}}
\caption{
MC events of class G: a) Relation between $\epts$ and $\egamma$;
b) Relation between $\egrid$ and $\egamma$.
The red circles are the average $\epts$ in the true energy $\egamma$ bin.
}
\label{eptsgrid}
\end{center}
\end{figure*}

The rational behind the use of $\epts$ as energy estimator in measuring the GRID resolutions 
is the expectation that the resolution of $\epts$ is much
better than that of $\egrid$.
That can be verified for MC events of class G comparing the distributions of $\epts$ versus 
$\egamma$ as well as $\egrid$ versus $\egamma$ and verifying that the former distribution
is much narrower that the latter. 
Those distributions are shown in Fig.\ref{eptsgrid}a and Fig.\ref{eptsgrid}b, respectively, 
with the red markers representing the average $\epts$ in $\egamma$ bins \SI{50}{\MeV} wide.
%The poor linearity of the circles is due to the presence of a small but not negligible background of 
%measurements at low $\epts$ with large $\egamma$. These events are
%due to electrons emitting $\gamma$-rays with $\egamma$\SI{>50}{\MeV} missing the PTS with another electron 
%from the same bunch hitting the final section of the PTS resulting in a fake signal at 
%$\epts\approx$\SI{50}{\MeV} \citep{btfagile}.
It is apparent that the resolution of $\epts$ as an energy estimator of $\egamma$ is much 
better than that of $\egrid$ and therefore
$\epts$ provides an effective energy estimator for calibrating the GRID with real data under 
the mild assumption that the energy resolution of $\epts$ for data is not much worse than 
that for MC.

\section{Results}

As shown in Table~\ref{data-samples}, data have been collected for different values of 
$\Theta$ and $\Phi$. In principle all figures of merit presented in Sect.~\ref{merit} depend
on $\Theta$ and $\Phi$ but within the statistics collected no dependence on $\Phi$ is visible.
Therefore, in the following, the data collected for different values of $\Phi$ and the same 
value of $\Theta$ are grouped together, and the results are presented as dependent on $\Theta$ only.

\subsection{Energy Dispersion Probability from real and MC data }

The EDP defined in Eq.\ref{edp} can be obtained illuminating uniformly the 
GRID with a monochromatic beam of $\gamma$-rays with energy $\egamma$ and considering 
the resulting spectrum of $\egrid$. This procedure, repeated 
for an adequate number of energies $\egamma$, allows to build a matrix $(\egamma, \egrid)$ 
defining the EDP.
This can be done with MC simulations but, not being available monochromatic 
$\gamma$-ray beams of variable and known energy with sufficient size to guarantee uniform
illumination, cannot be done experimentally.

In any case, this approach is not the most appropriate for the application to which GRID
is dedicated. The $\gamma$-ray spectra measured by GRID during its operation in flight
are power-law spectra like $1/\egamma^\alpha$ with $\alpha\sim 1.7$ \citep{chenaa}, 
therefore a matrix is built with MC events produced according to this
power-law spectrum $(\egamma, \egrid)^\mathrm{MC}_{1.7}$ (the subscript is the power index).

The goal of the BTF calibration is to assess quantitatively the reliability of 
the MC simulation used to build the EDP matrix, not to measure the EDP matrix directly.
One reason is that the BTF $\gamma$-ray beam follows an approximate $1/\egamma$ 
Bremsstrahlung spectrum on the restricted energy range available producing an EDP 
matrix different from the one estimated using a power-law spectrum whose spectral 
index is $\alpha = −1.7$ \citep{chenaa}; the other reason is that the true energy 
$\egamma$ of the single $\gamma$-ray is unknown.
At the BTF, a straightforward comparison is possible between 
$(\epts, \egrid)^\mathrm{data}_{1.0}$ and $(\epts, \egrid)^\mathrm{MC}_{1.0}$.
Our goal is to estimate the effect of the discrepancies measured in this comparison of the
errors between $(\egamma, \egrid)^\mathrm{data}_{1.7}$ and $(\egamma, \egrid)^\mathrm{MC}_{1.7}$,
that are the actual systematic errors in the EDP.

\begin{figure*}[htb]
\begin{center} 
\mbox{\begin{tabular}[t]{cc}
\subfigure[]{\includegraphics[width=0.48\textwidth]{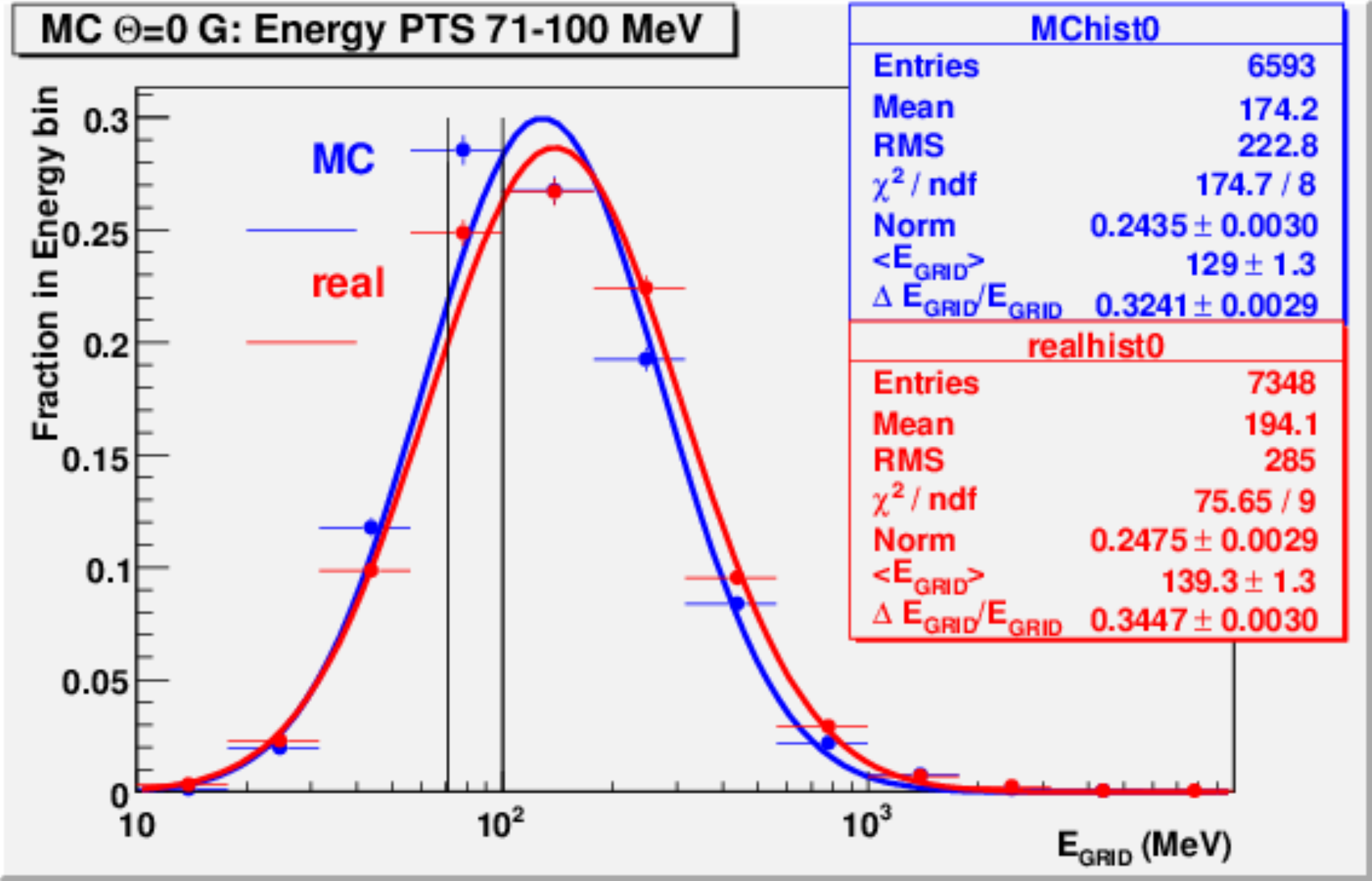} } &
\subfigure[]{\includegraphics[width=0.48\textwidth]{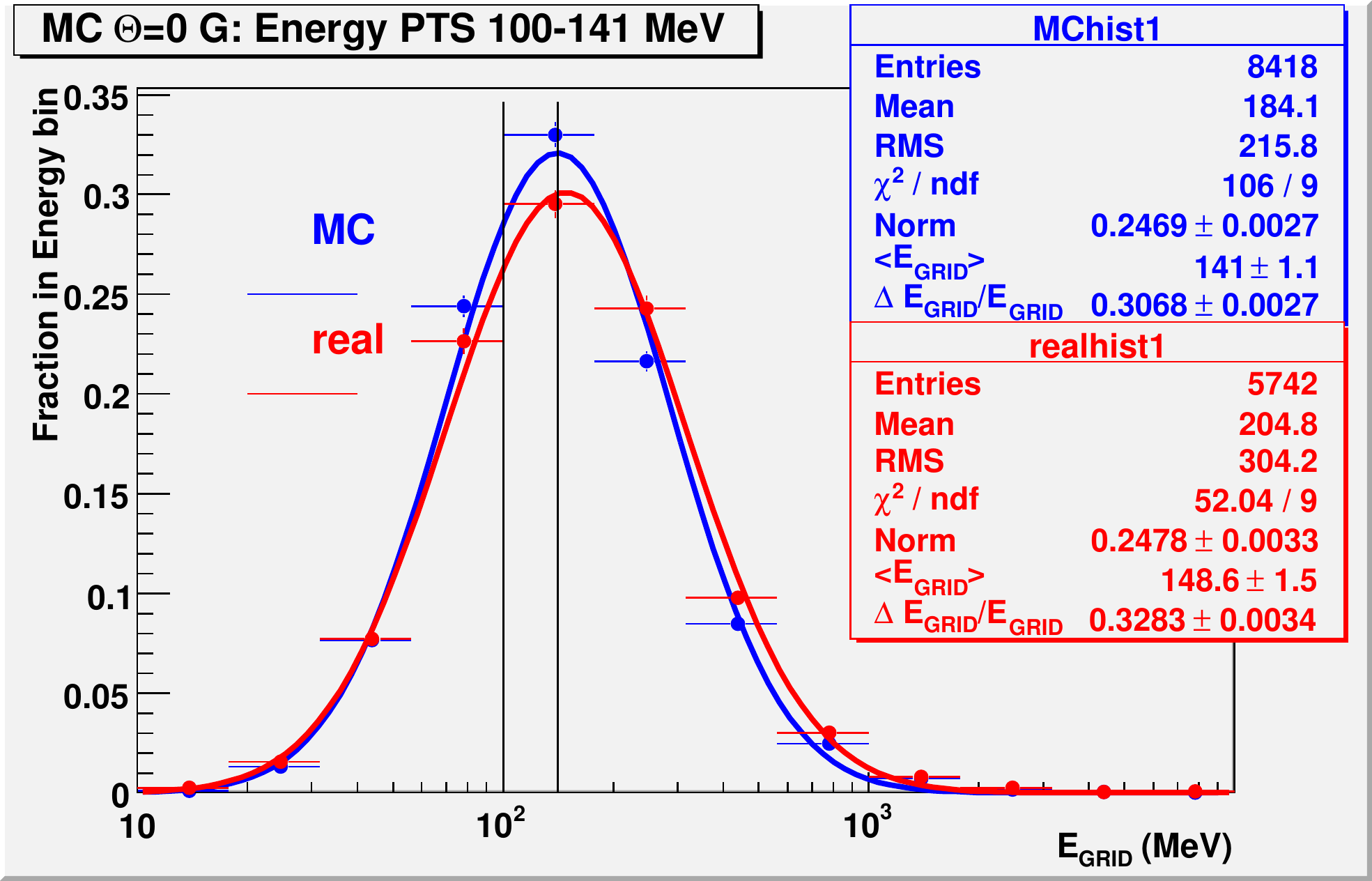} } \\
\subfigure[]{\includegraphics[width=0.48\textwidth]{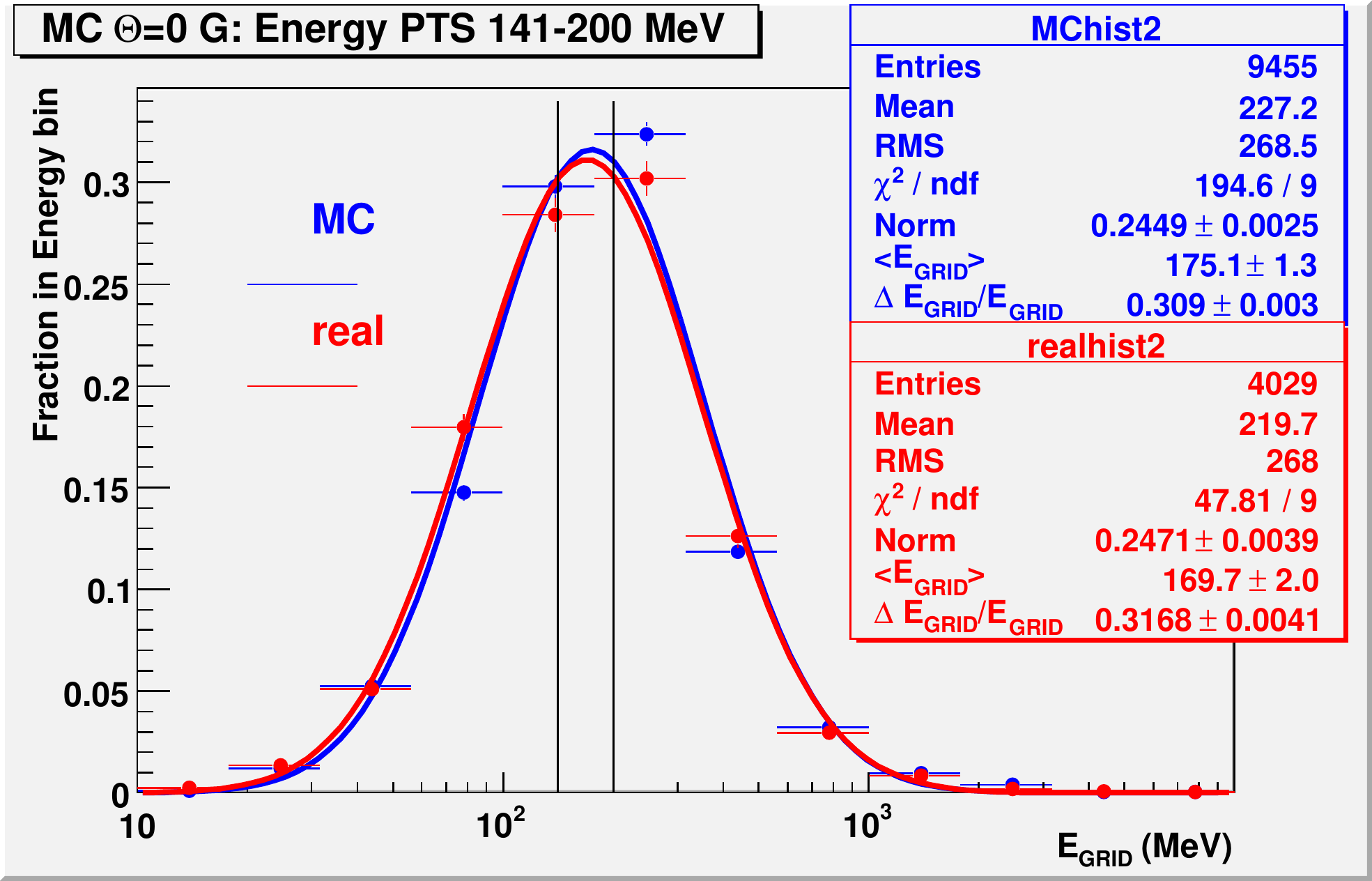} } &
\subfigure[]{\includegraphics[width=0.48\textwidth]{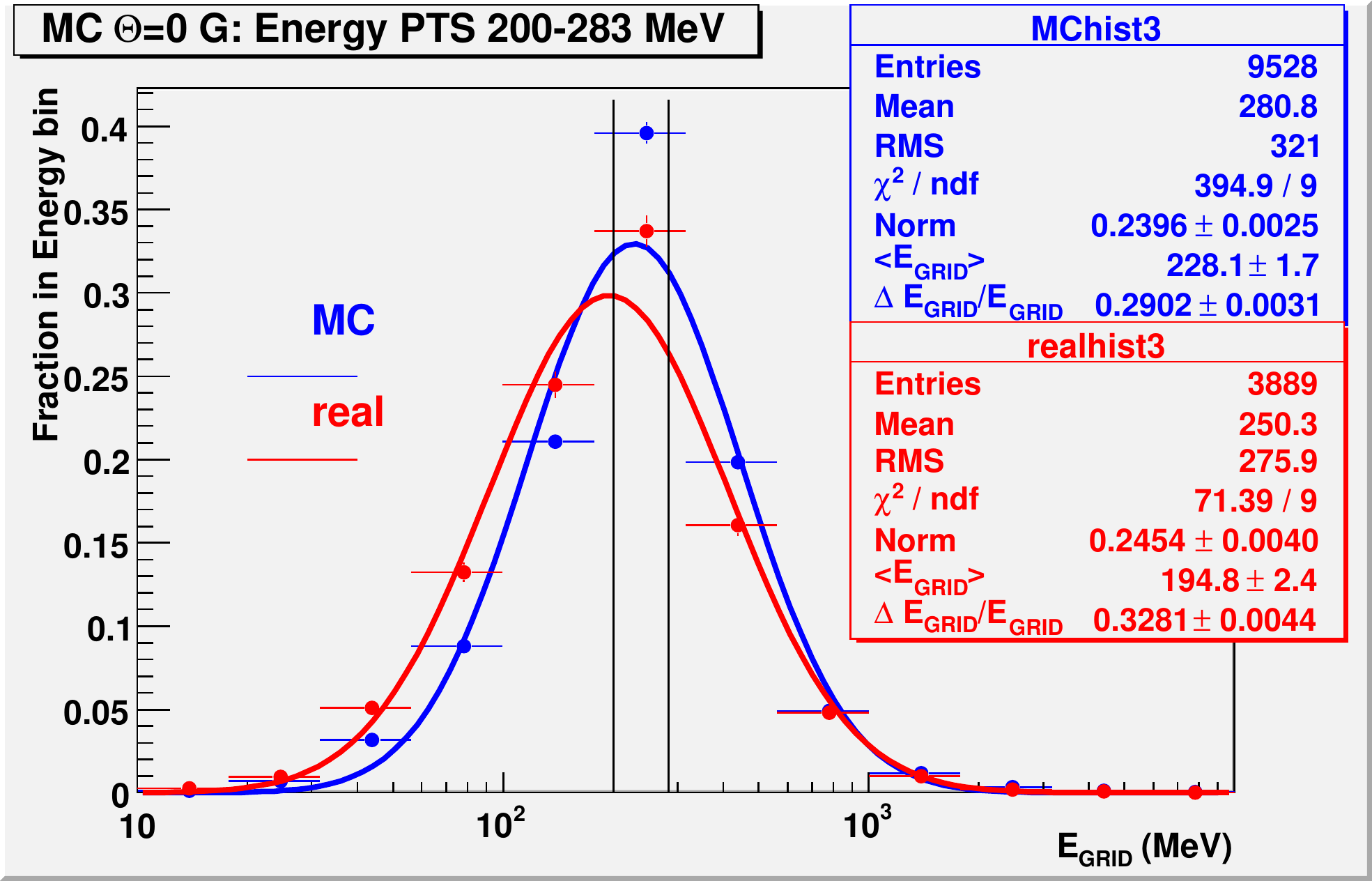} } \\
\subfigure[]{\includegraphics[width=0.48\textwidth]{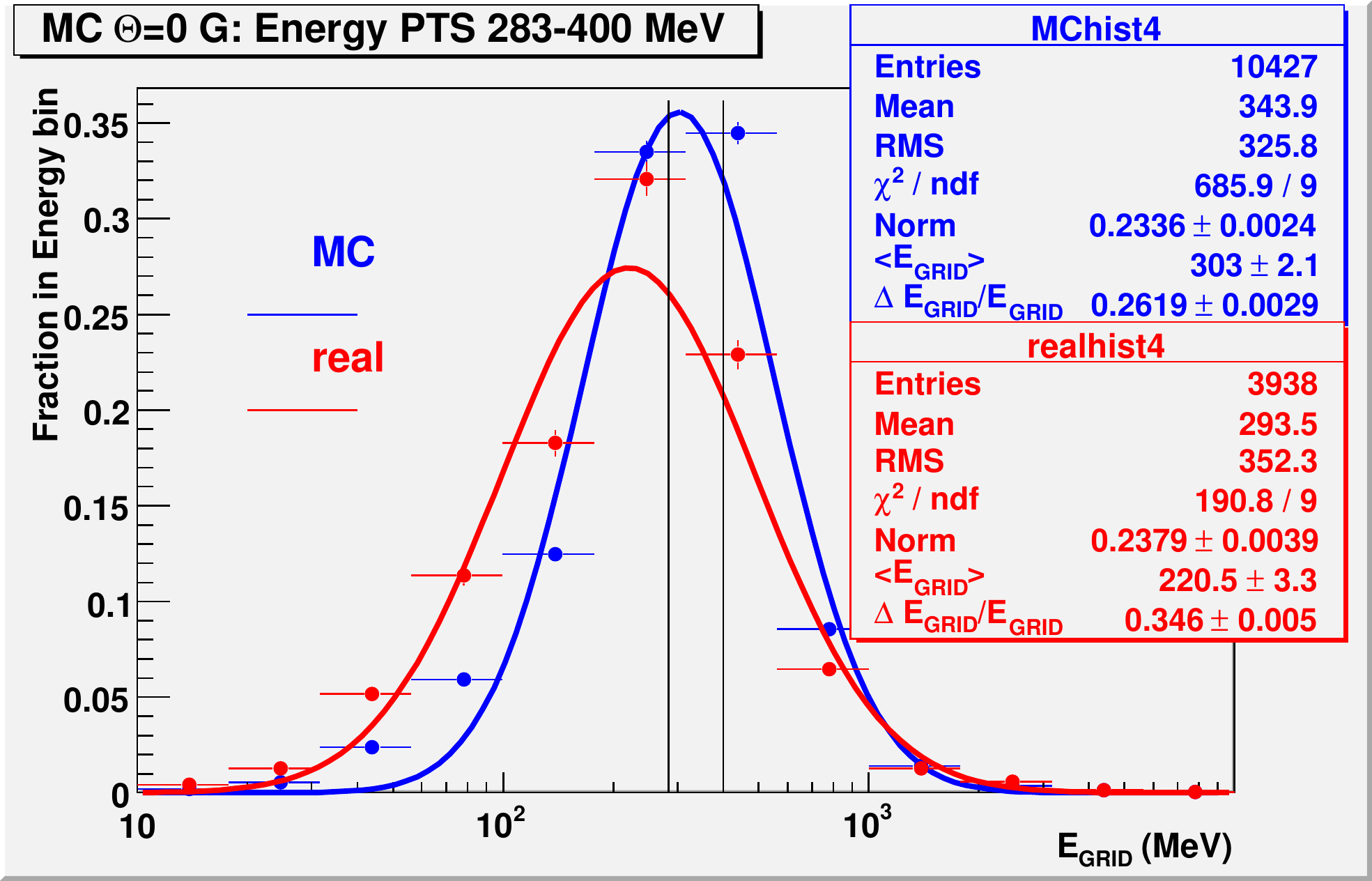} } \\
\end{tabular}}
\caption{Energy Dispersion Probability for MC (blue) and real data (red) in five bins of 
$\epts$ for events of class G and $\Theta$=\ang{0} for 
a) \SI{71}{\MeV}  $\le \epts \le$ \SI{100}{\MeV} 
b) \SI{100}{\MeV} $\le \epts \le$ \SI{141}{\MeV} 
c) \SI{141}{\MeV} $\le \epts \le$ \SI{200}{\MeV} 
d) \SI{200}{\MeV} $\le \epts \le$ \SI{283}{\MeV} 
e) \SI{283}{\MeV} $\le \epts \le$ \SI{400}{\MeV}.
The pair of vertical lines delimit the $\epts$ range used to produce each histograms.
See text for more detail.
}
\label{psfener}
\end{center}
\end{figure*}

In Fig.\ref{psfener} the events are partitioned in five $\epts$ bins and 
the distributions of $\egrid$ are displayed for MC (blue) and real data (red).
The variable $\log(\egrid)$ is fitted with a Gaussian separately for MC and data in each $\epts$ bin.
The fitted averages and the relative standard deviations are reported in the legend 
as $\langle\egrid\rangle$ and $\Delta \egrid/\langle\egrid\rangle$.

The values of $\langle\egrid\rangle$ for each energy bin are within or close to the 
$\epts$ bins both for MC and real data with the exception of Fig.\ref{psfener}a, where,
as visible in Fig.~\ref{eptsgrid}a, there is in the low energy $\epts$ bin a significant tail
due to high $\egrid$ events. 
The widths of the distributions for MC and data are well compatible with each other within 
5\%, that can be assumed as an upper limit estimation of the systematic error,
with some discrepancy in Fig.\ref{psfener}e for $\epts$ \SI{\approx 400}{\MeV}, and consistent 
with the value estimated in Sect.~\ref{enermeas}. 
The discrepancy in Fig.\ref{psfener}e may be due to the poor reliability of $\epts$ 
as estimator of $\egamma$ at this energy, as explained in Sect.~\ref{ptsenerest}.

Similar results are obtained for different values of $\Theta$.

\begin{figure*}[htb]
\begin{center}
\mbox{\begin{tabular}[t]{cc}
\subfigure[]{\includegraphics[width=0.48\textwidth]{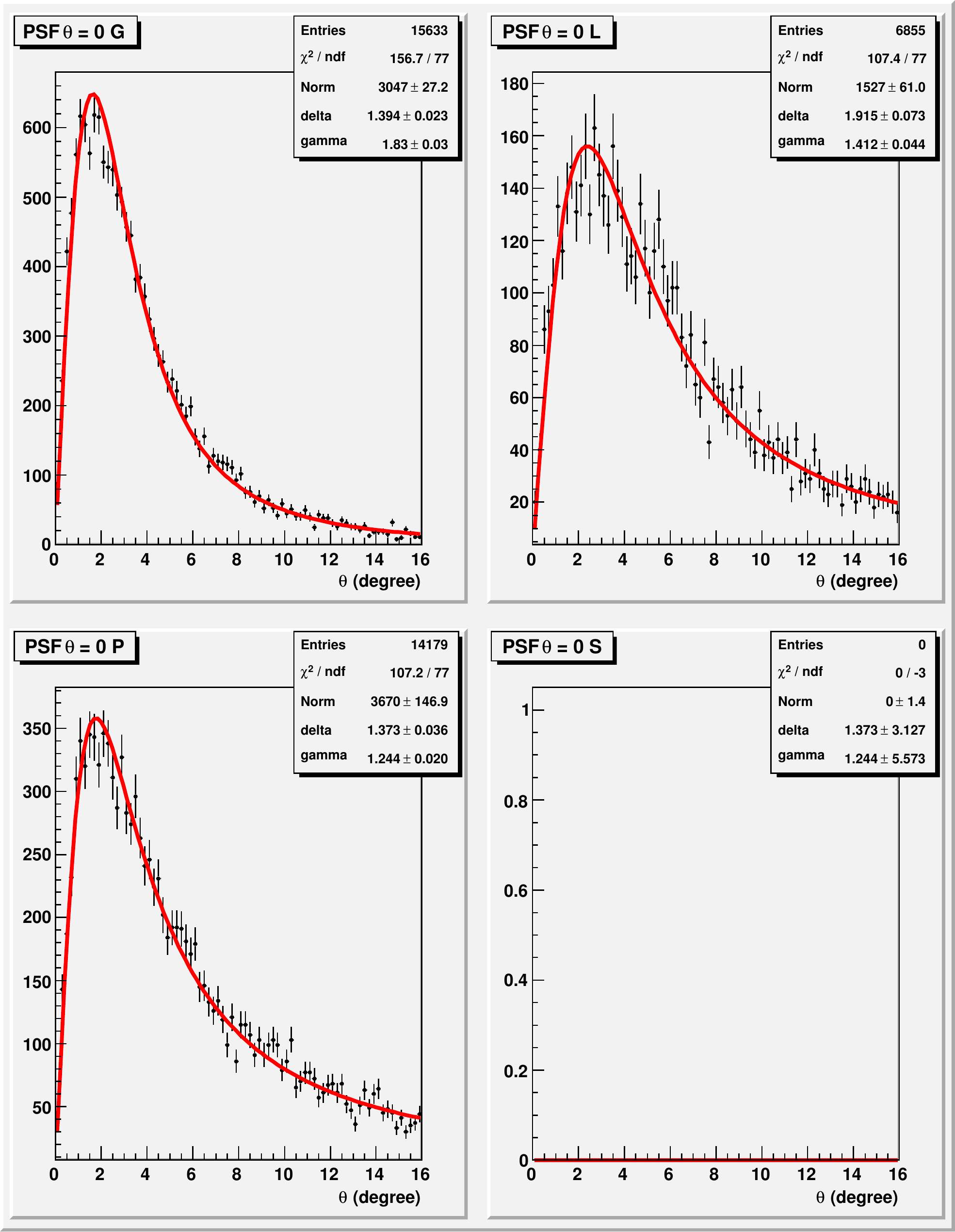}
} &
\subfigure[]{\includegraphics[width=0.48\textwidth]{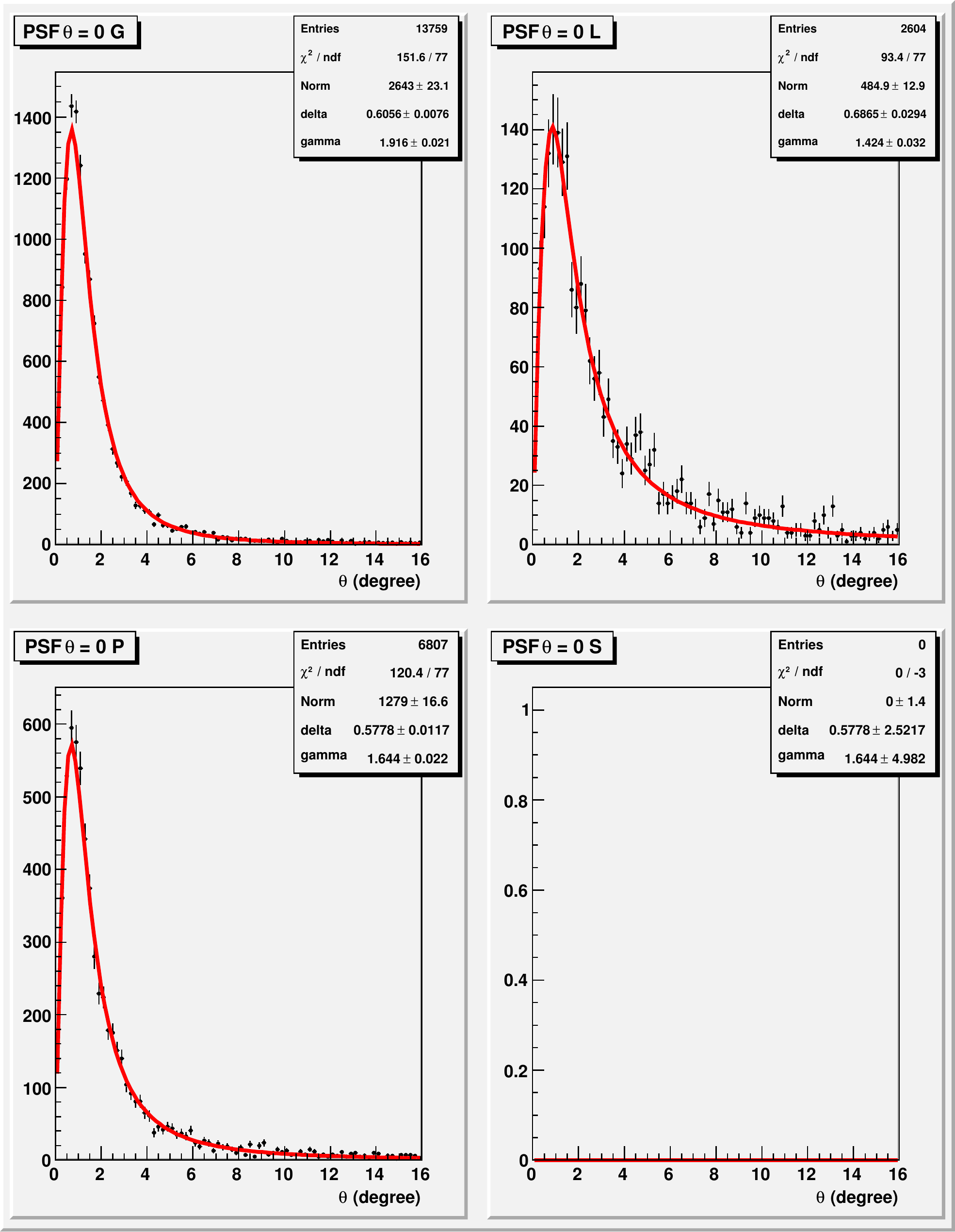}
} \\
\subfigure[]{\includegraphics[width=0.48\textwidth]{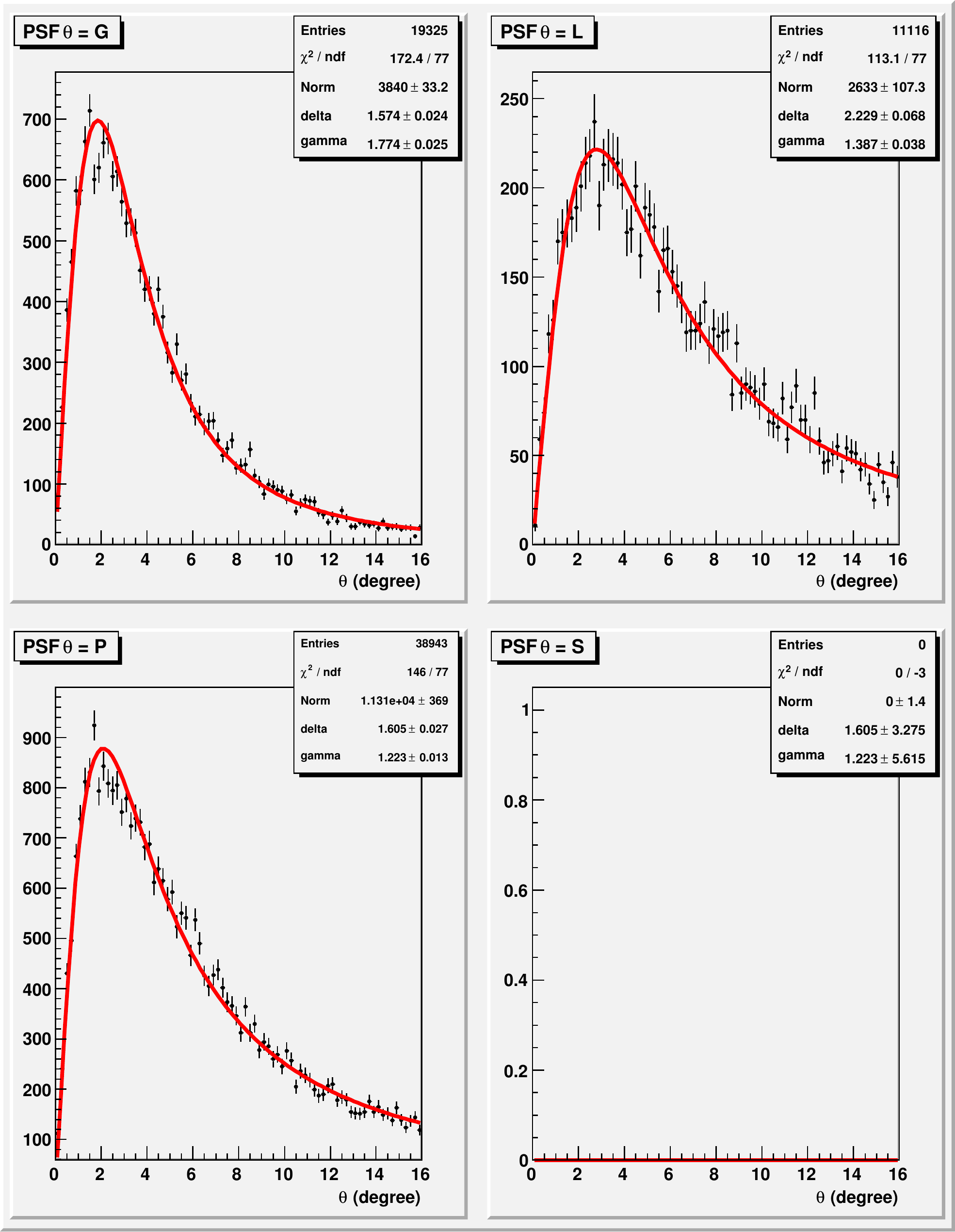}
} &
\subfigure[]{\includegraphics[width=0.48\textwidth]{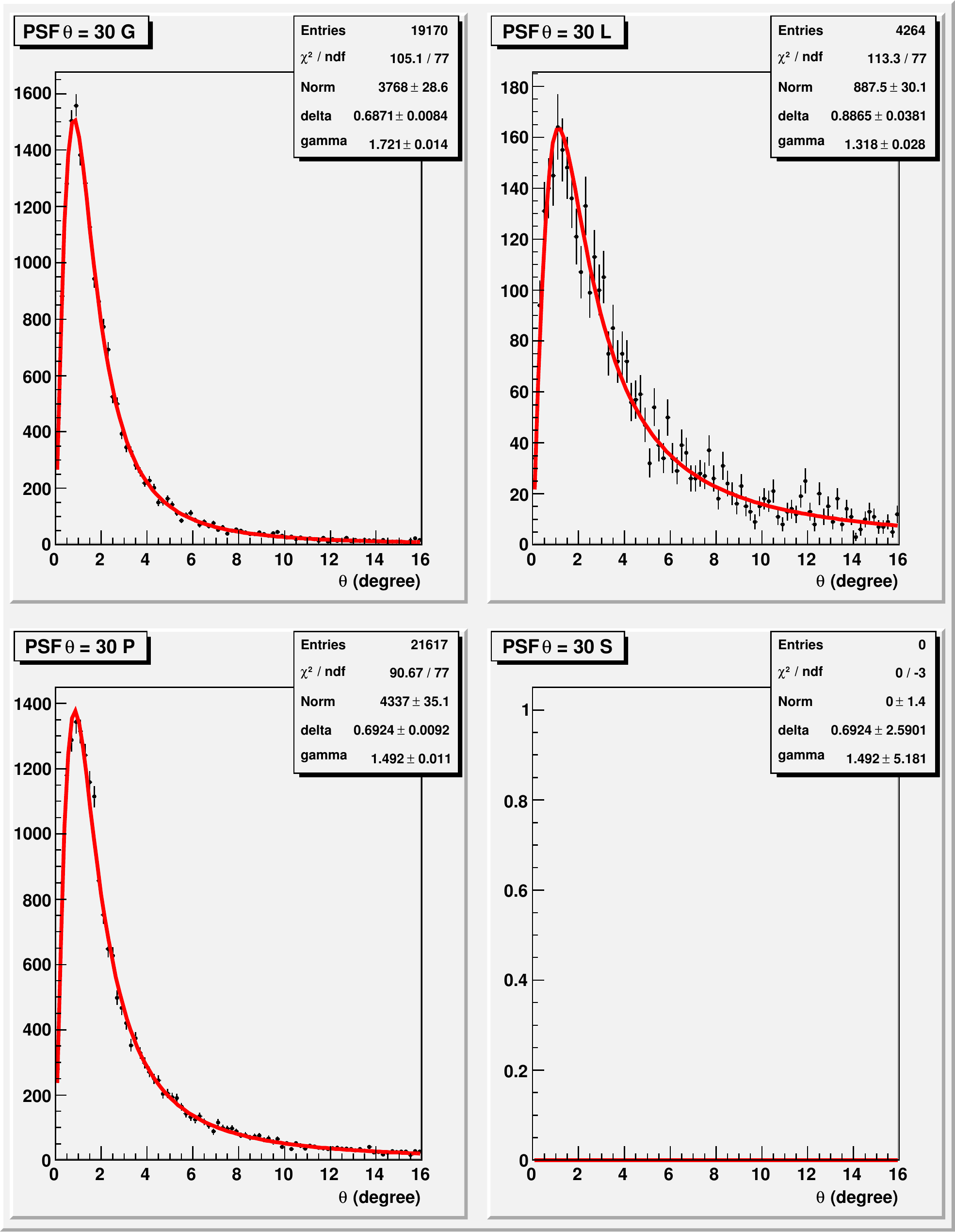}
} \\
\end{tabular}}
\caption{Distributions of the three-dimensional angular difference for MC events with the PSF fitted with King function 
(the four plots in each quadrant corresponds to events of class G,L,P and S respectively) for 
a) $\Theta$=\ang{0}, \SI{70}{\MeV} $\le \egrid \le$ \SI{140}{\MeV} 
b) $\Theta$=\ang{0}, \SI{280}{\MeV} $\le \egrid \le$ \SI{420}{\MeV} 
c) $\Theta$=\ang{30}, \SI{70}{\MeV} $\le \egrid \le$ \SI{140}{\MeV} 
d) $\Theta$=\ang{30}, \SI{280}{\MeV} $\le \egrid \le$ \SI{420}{\MeV}.
}
\label{psfkingmc}
\end{center}
\end{figure*}

\begin{figure*}[htb]
\begin{center}
\mbox{\begin{tabular}[t]{cc}
\subfigure[]{\includegraphics[width=0.48\textwidth]{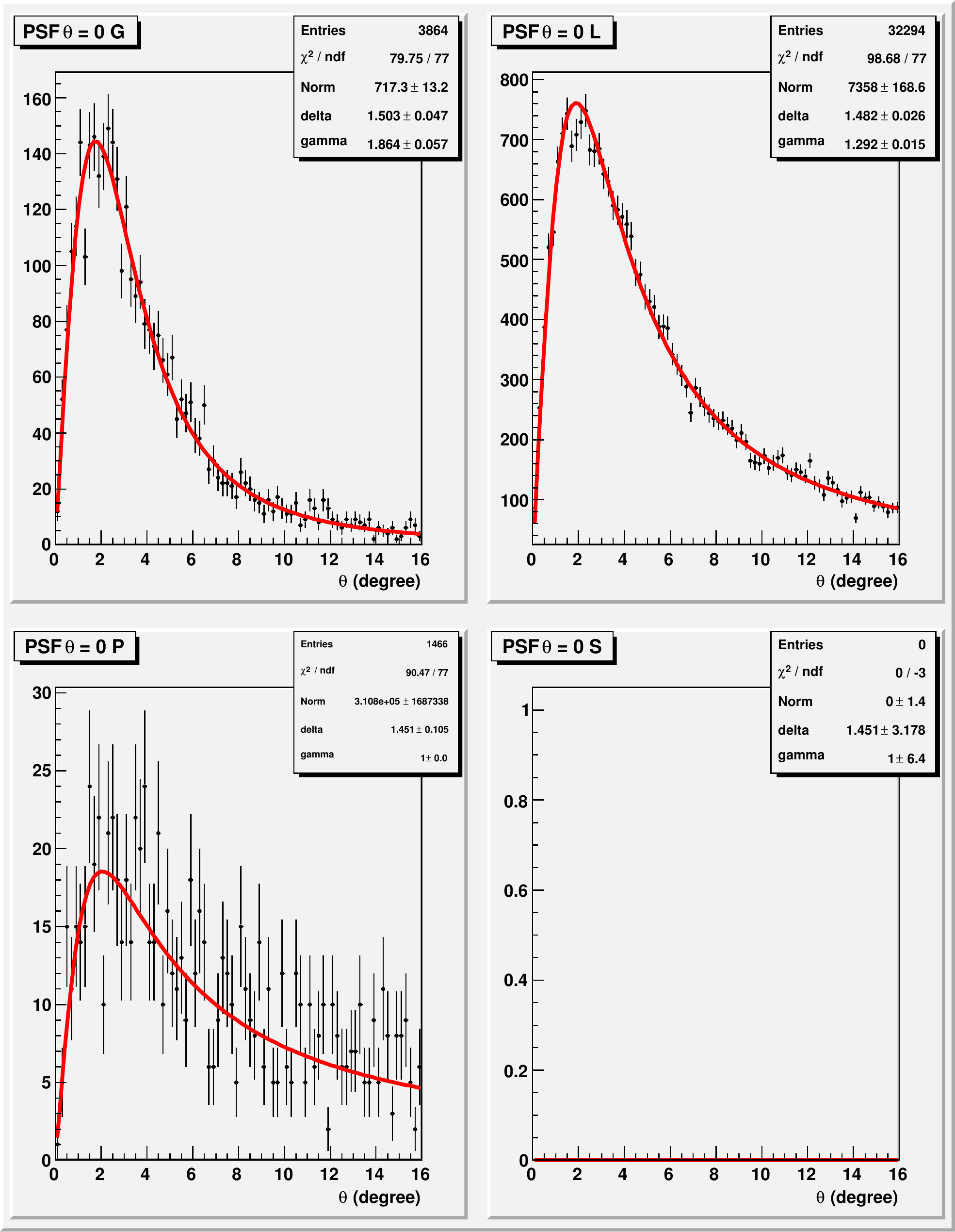}
} &
\subfigure[]{\includegraphics[width=0.48\textwidth]{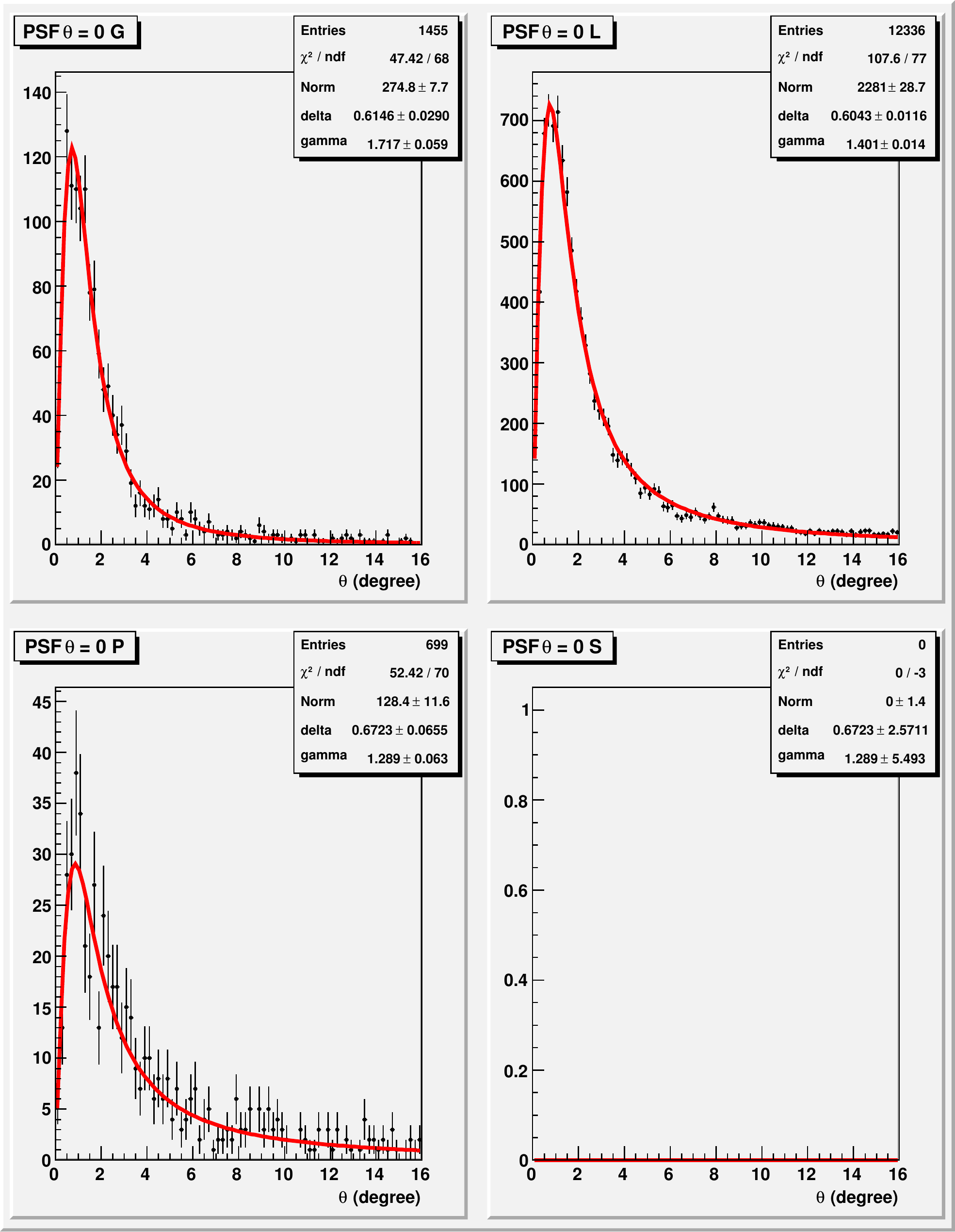}
} \\
\subfigure[]{\includegraphics[width=0.48\textwidth]{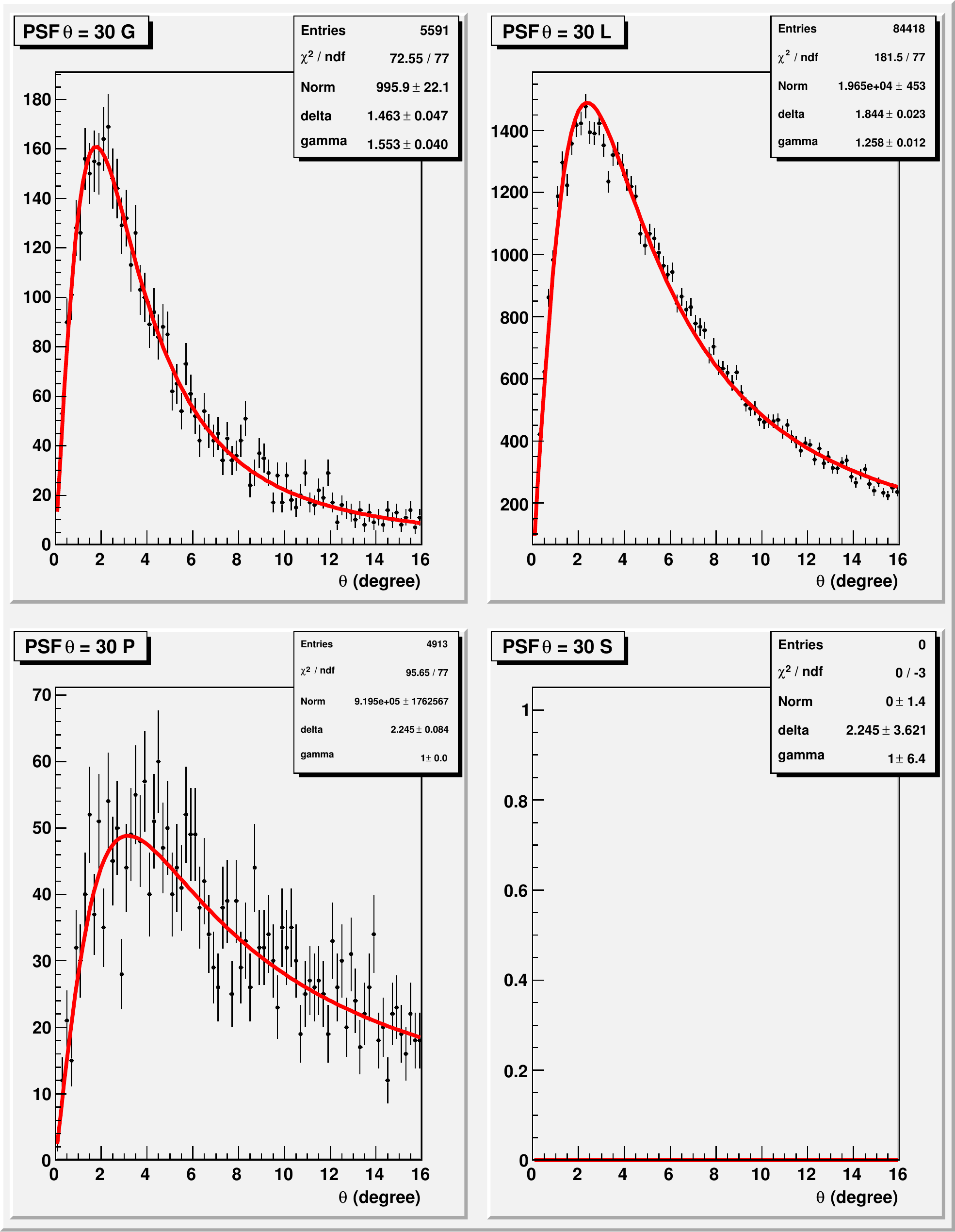}
} &
\subfigure[]{\includegraphics[width=0.48\textwidth]{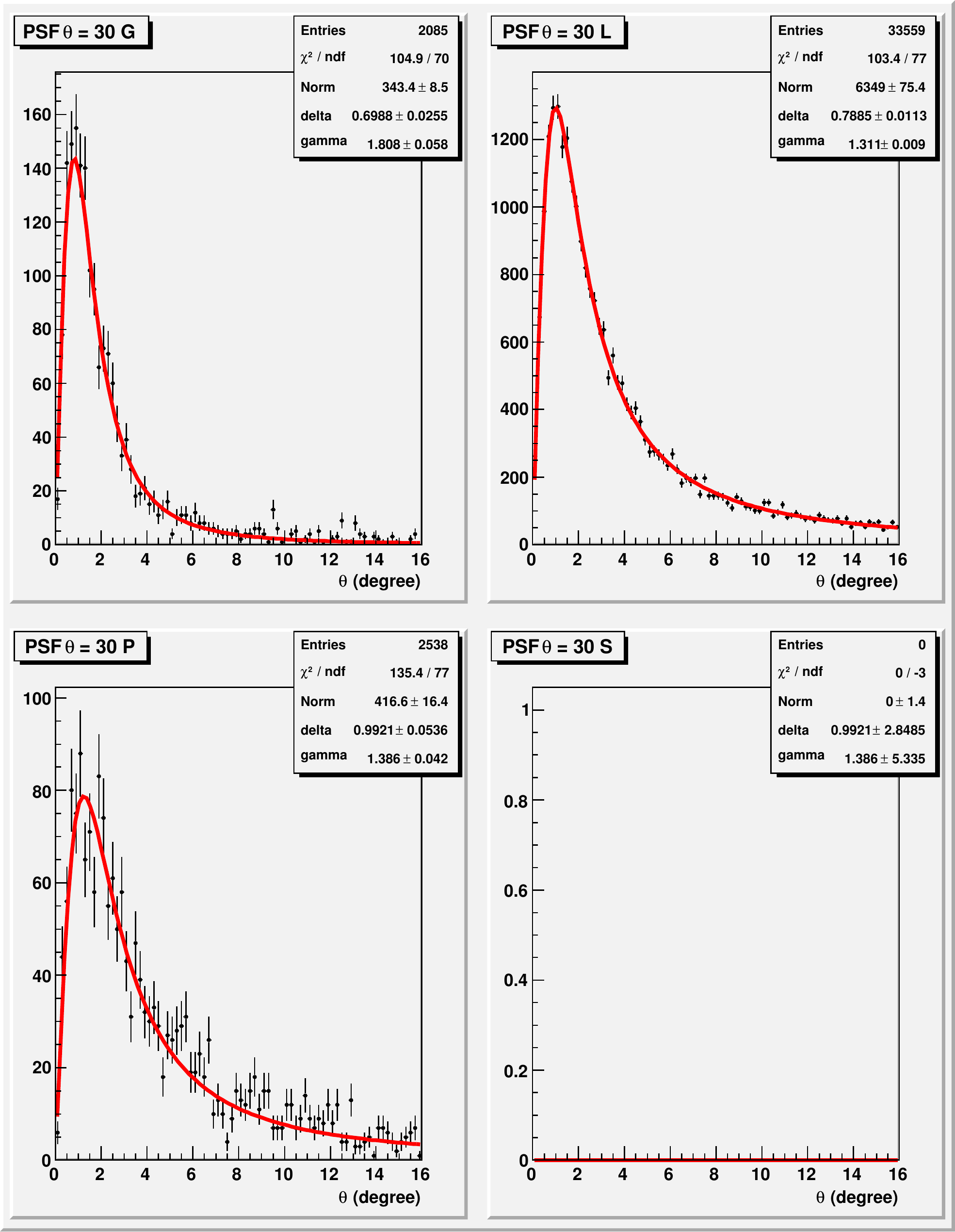}
} \\
\end{tabular}}
\caption{Distributions of the three-dimensional angular difference for real events with the PSF fitted with King function 
(the four plots in each quadrant corresponds to events of class G,L,P and S respectively) for 
a) $\Theta$=\ang{0}, \SI{70}{\MeV} $\le \egrid \le$ \SI{140}{\MeV} 
b) $\Theta$=\ang{0}, \SI{280}{\MeV} $\le \egrid \le$ \SI{420}{\MeV} 
c) $\Theta$=\ang{30}, \SI{70}{\MeV} $\le \egrid \le$ \SI{140}{\MeV} 
d) $\Theta$=\ang{30}, \SI{280}{\MeV} $\le \egrid \le$ \SI{420}{\MeV}.
}
\label{psfkingreal}
\end{center}
\end{figure*}

\subsection{PSF from real and MC data }

The performance of the instrument with respect to the angular resolution is 
characterised by the Point Spread Function (PSF) as defined in Sect.\ref{sec:intro}.
The distribution of the three-dimensional angular difference is fitted assuming 
that the PSF is described by the two parameters King's function as defined in the Appendix.
In addition to the function parameters, $\delta$ related to the standard deviation
and $\gamma$ related to non-Gaussian tail, also the value of the Containment Radius at 
68.3\%\ ($CR_{68}$) is quoted: this value is determined by the function parameters,
but it can be defined independently from the parametrisation.

\begin{figure*}[htb]
\begin{center}
\mbox{\begin{tabular}[t]{ccc}
\subfigure[]{\includegraphics[width=0.30\textwidth]{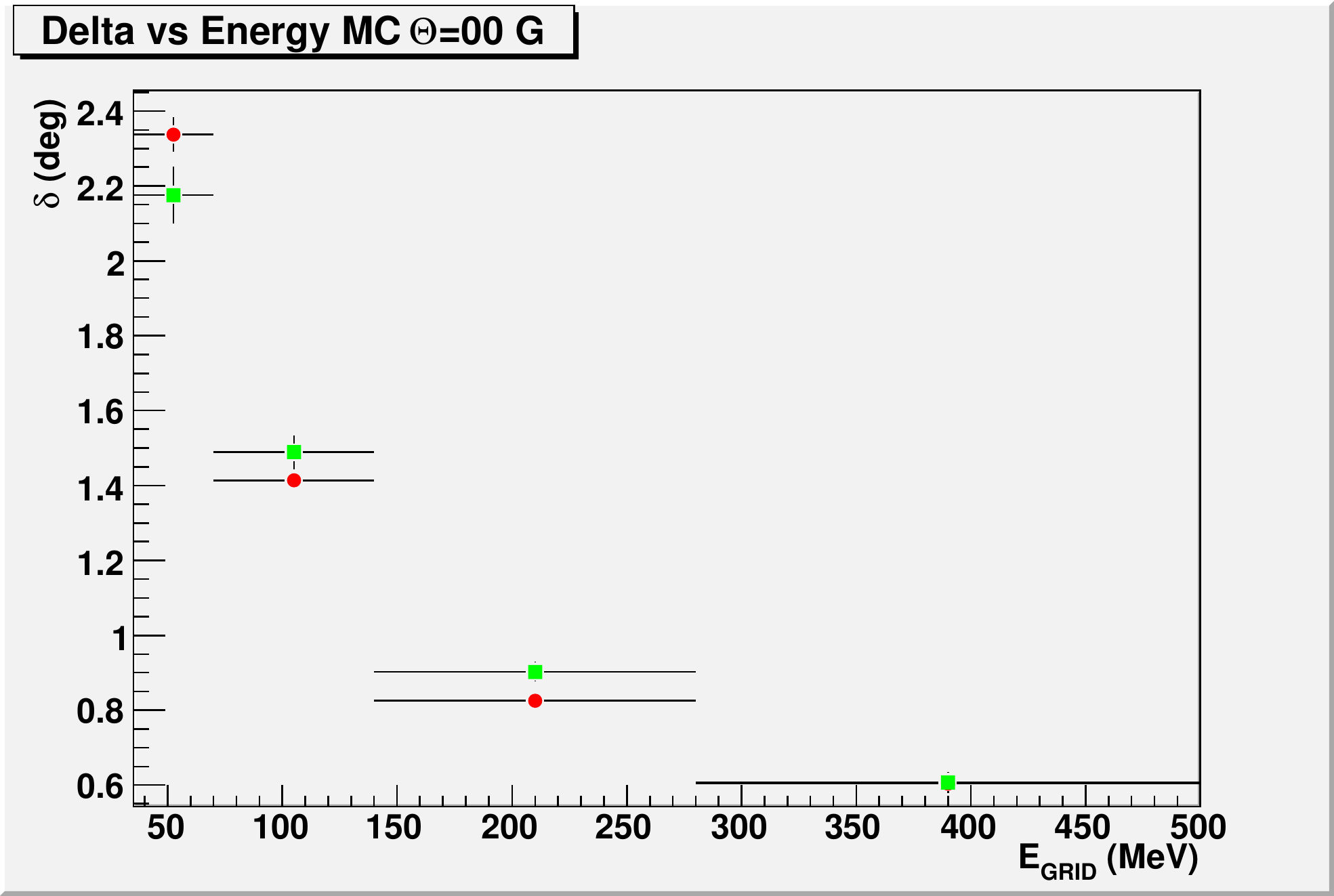}} &

\subfigure[]{\includegraphics[width=0.30\textwidth]{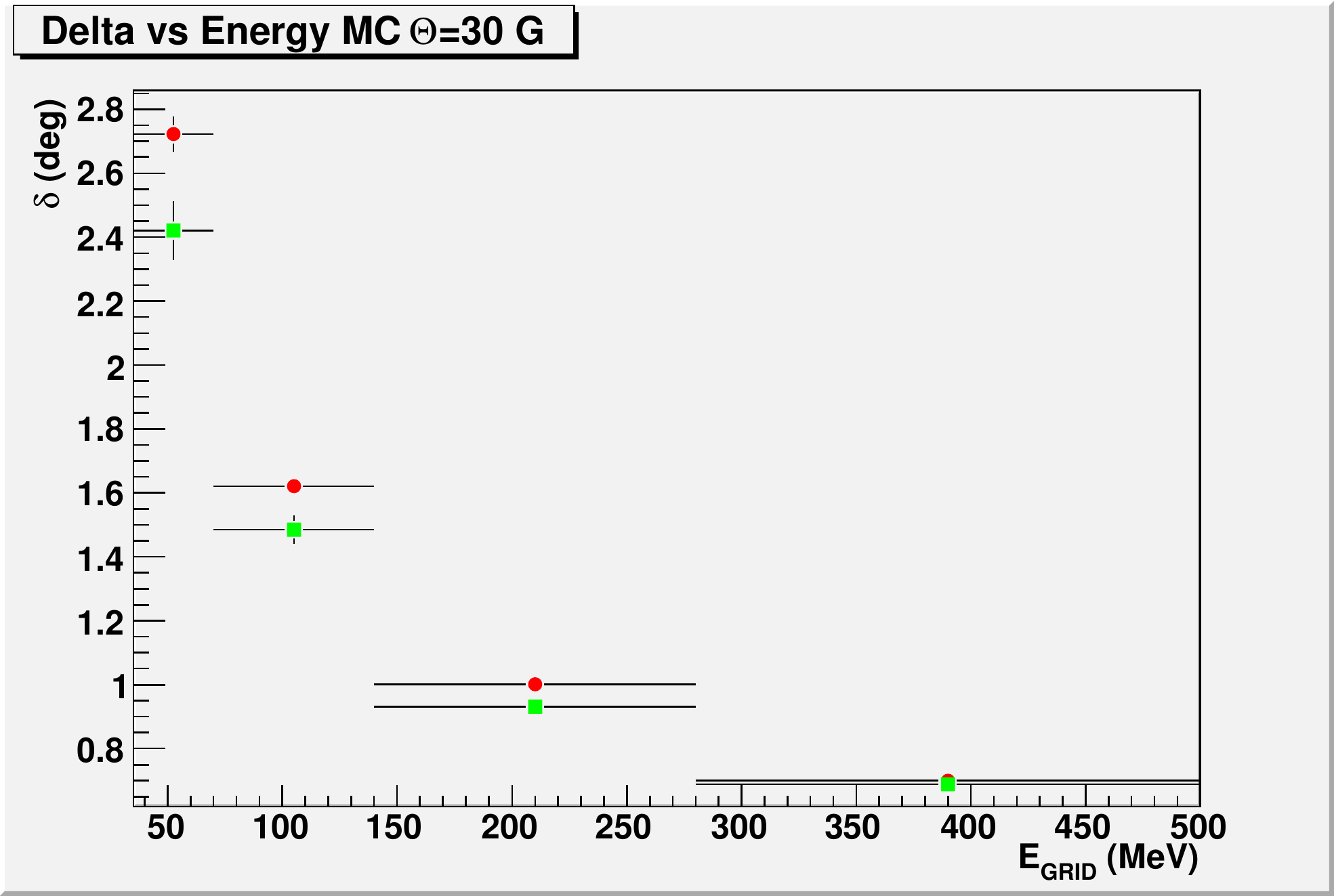}} &
  
\subfigure[]{\includegraphics[width=0.30\textwidth]{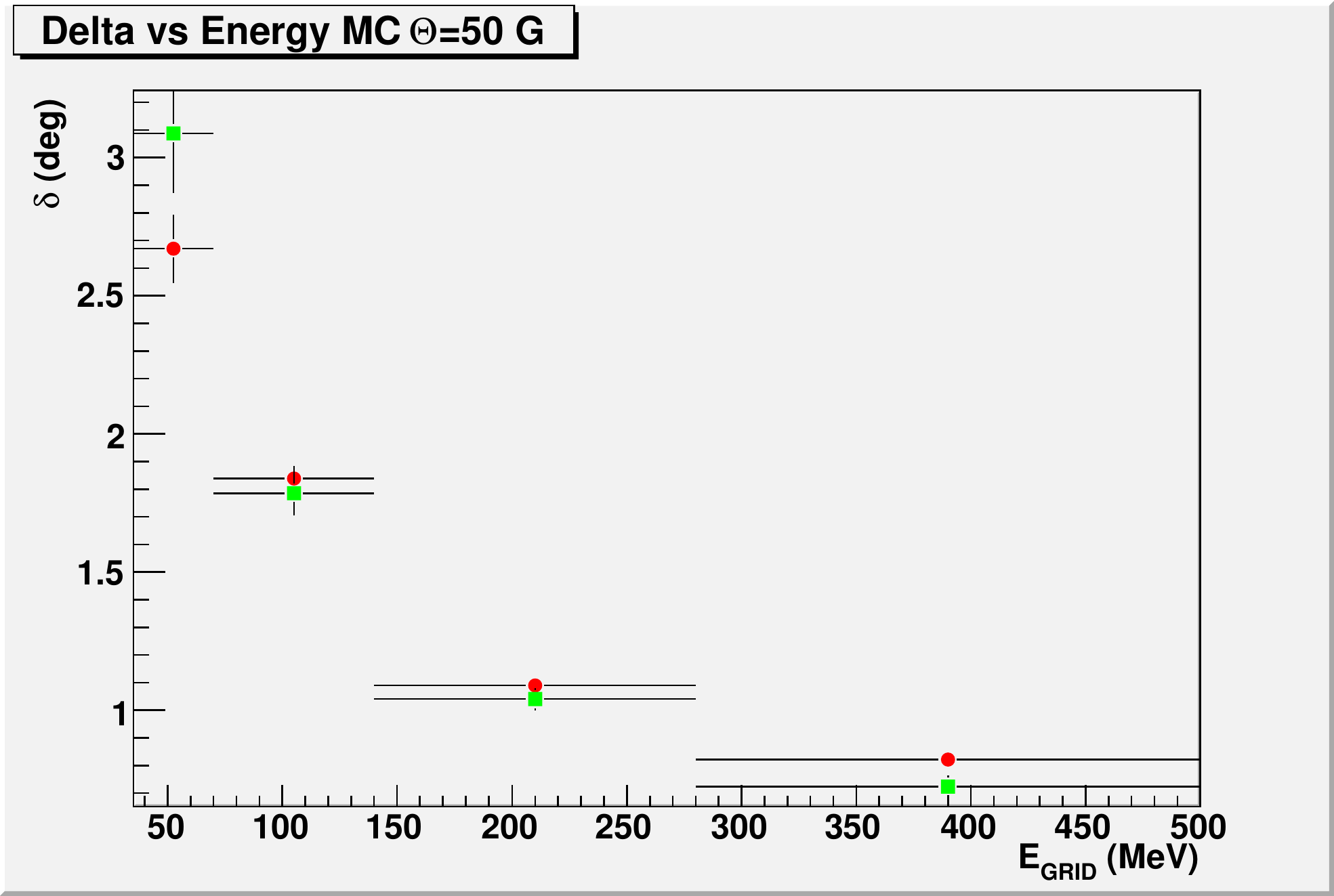}} 
\\
\end{tabular}}
\caption{Parameter $\delta$ of the King's function versus measured $\gamma$-ray energy $\egrid$ for real (green squares) and MC (red circles) events:
a) $\Theta$=\ang{0}, b) $\Theta$=\ang{30}, c) $\Theta$=\ang{50} 
}
\label{demcreal}
\end{center}
\end{figure*}

\begin{figure*}[htb]
\begin{center}
\mbox{\begin{tabular}[t]{ccc}
\subfigure[]{\includegraphics[width=0.3\textwidth]{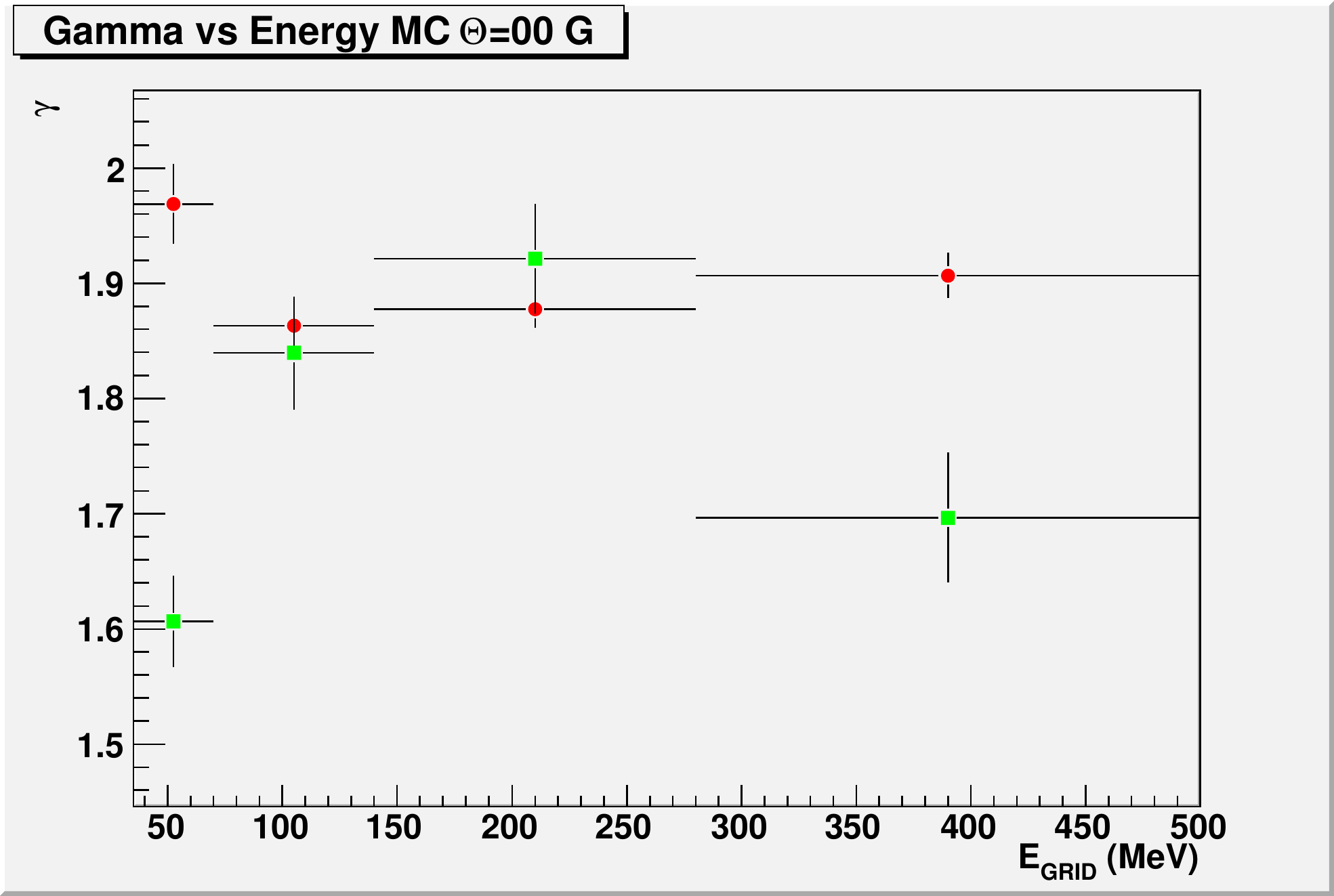}} &

\subfigure[]{\includegraphics[width=0.3\textwidth]{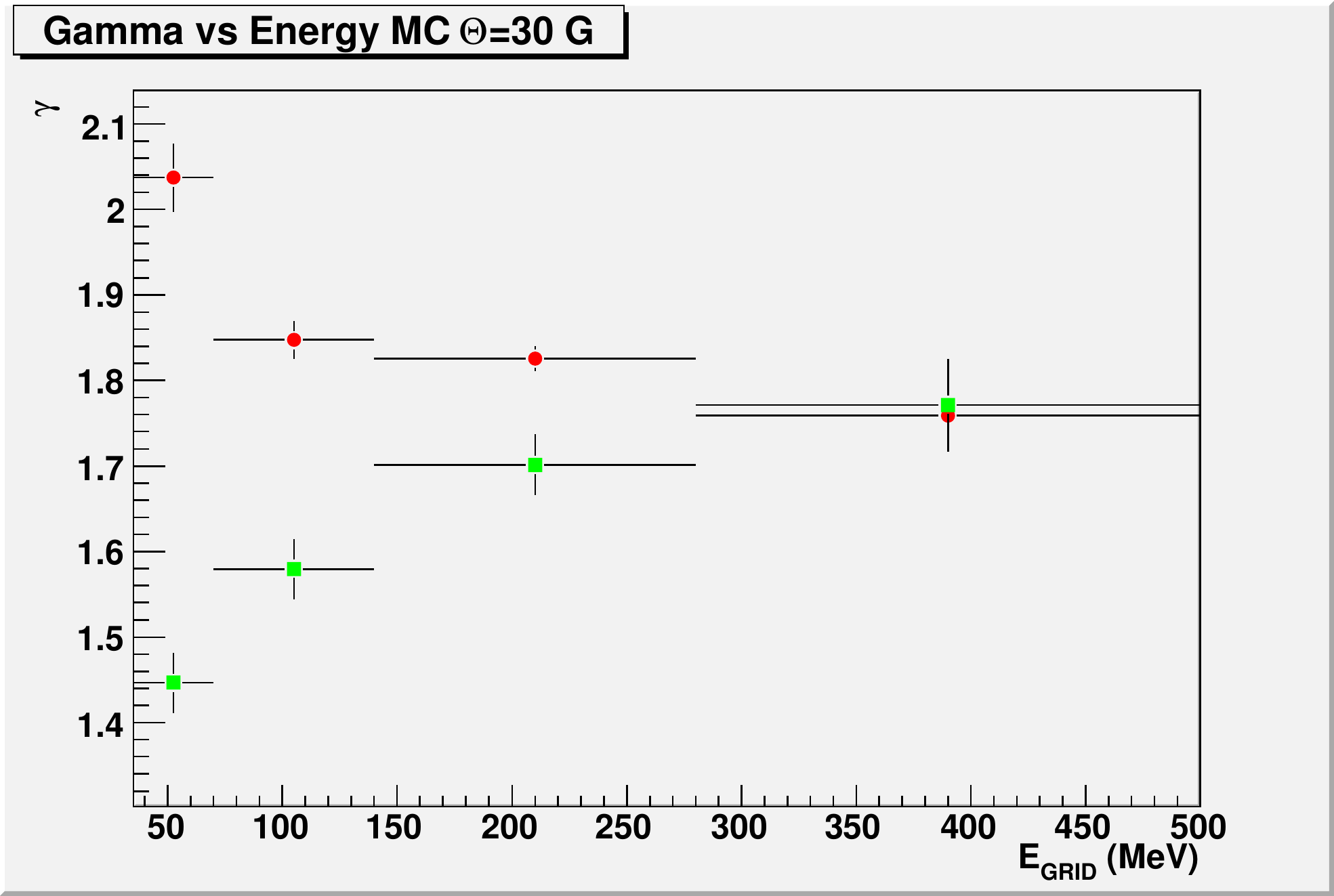}} &

\subfigure[]{\includegraphics[width=0.3\textwidth]{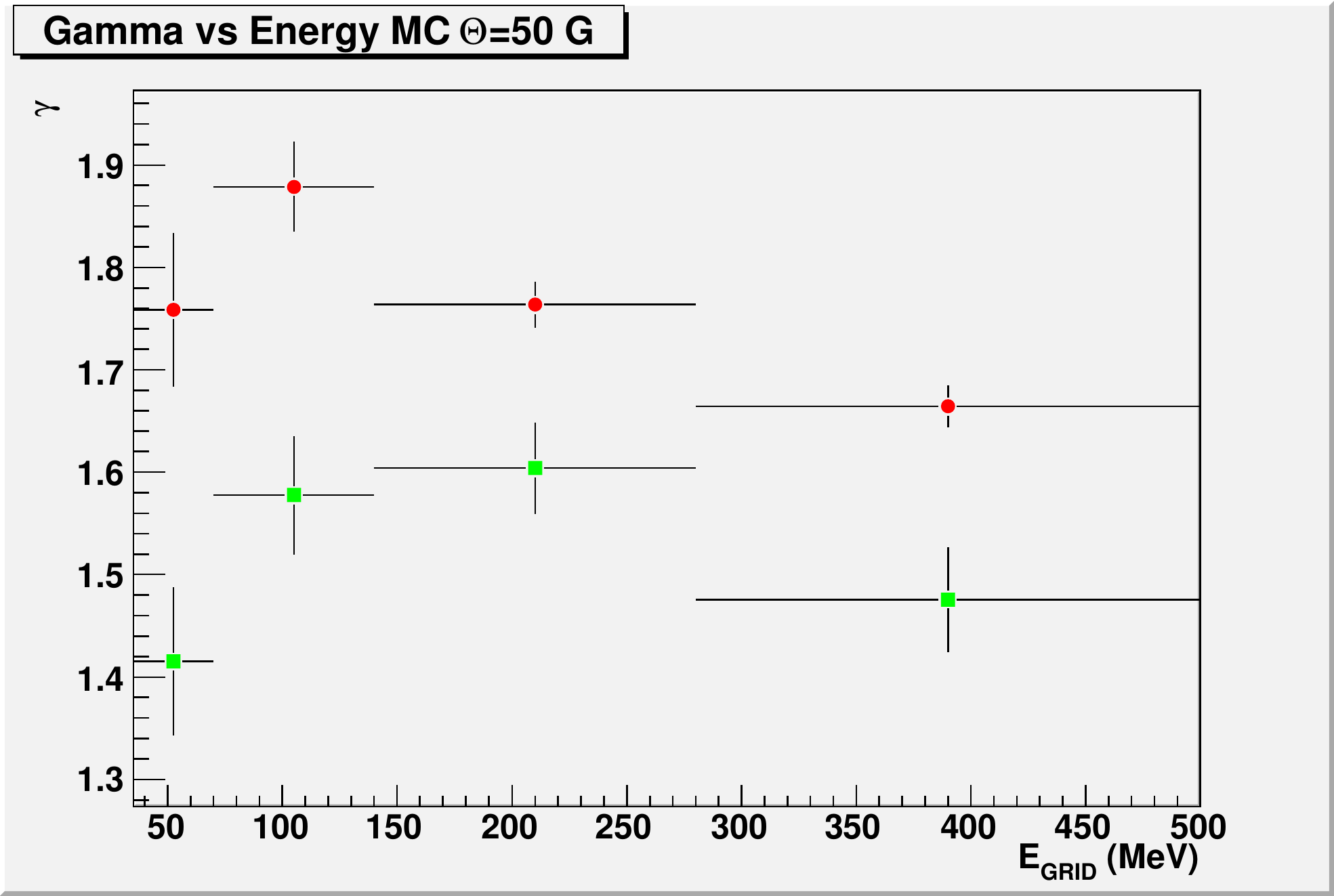}} 
\\
\end{tabular}}
\caption{Parameter $\gamma$ of the King's function measured versus $\gamma$-ray energy $\egrid$ for real (green squares) and MC (red circles) events 
a) $\Theta=\ang{0}$, b) $\Theta=\ang{30}$, c) $\Theta=\ang{50}$ 
}
\label{gemcreal}
\end{center}
\end{figure*}

\begin{figure*}[htb]
\begin{center}
\mbox{\begin{tabular}[t]{ccc}
\subfigure[]{\includegraphics[width=0.3\textwidth]{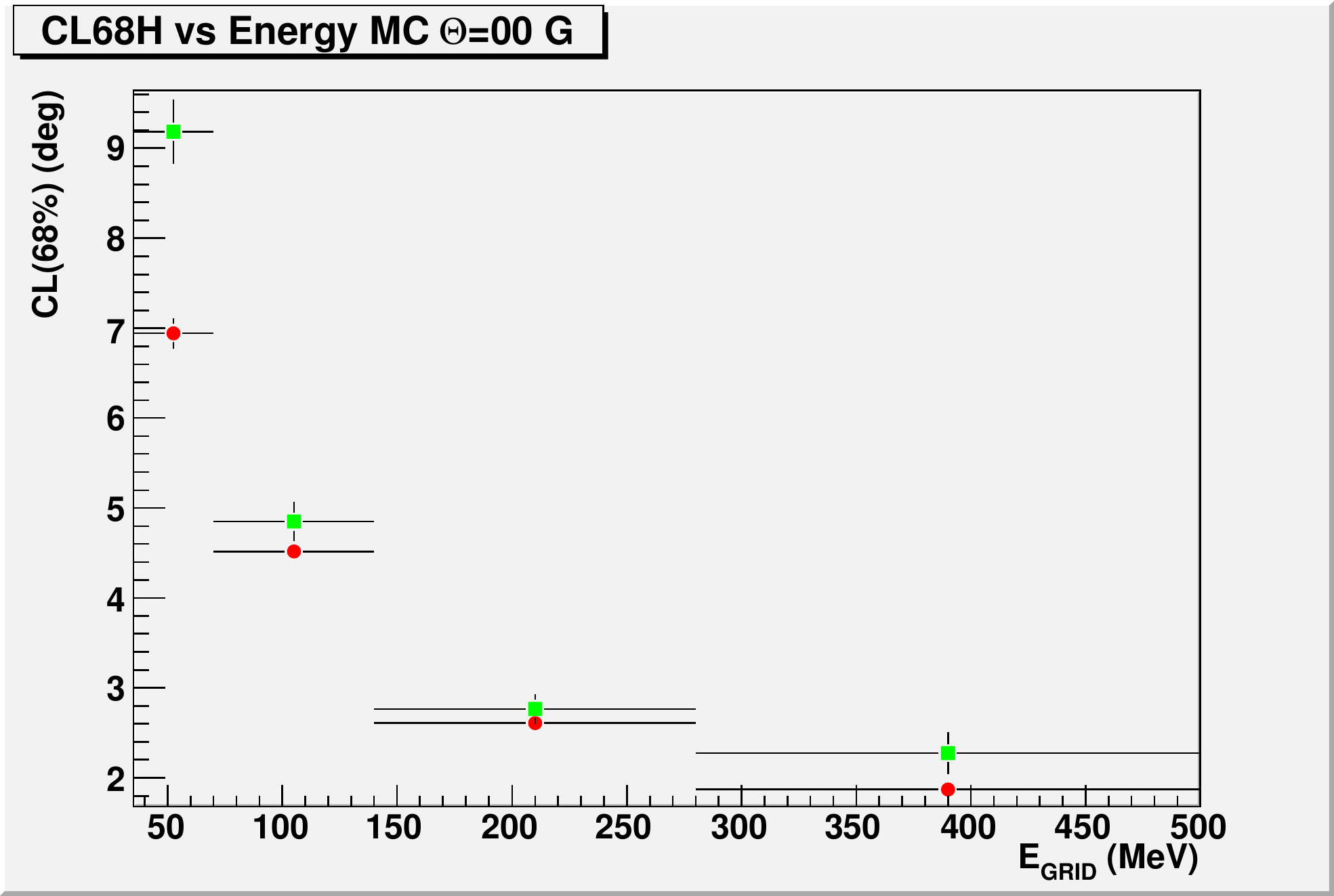}} &

\subfigure[]{\includegraphics[width=0.3\textwidth]{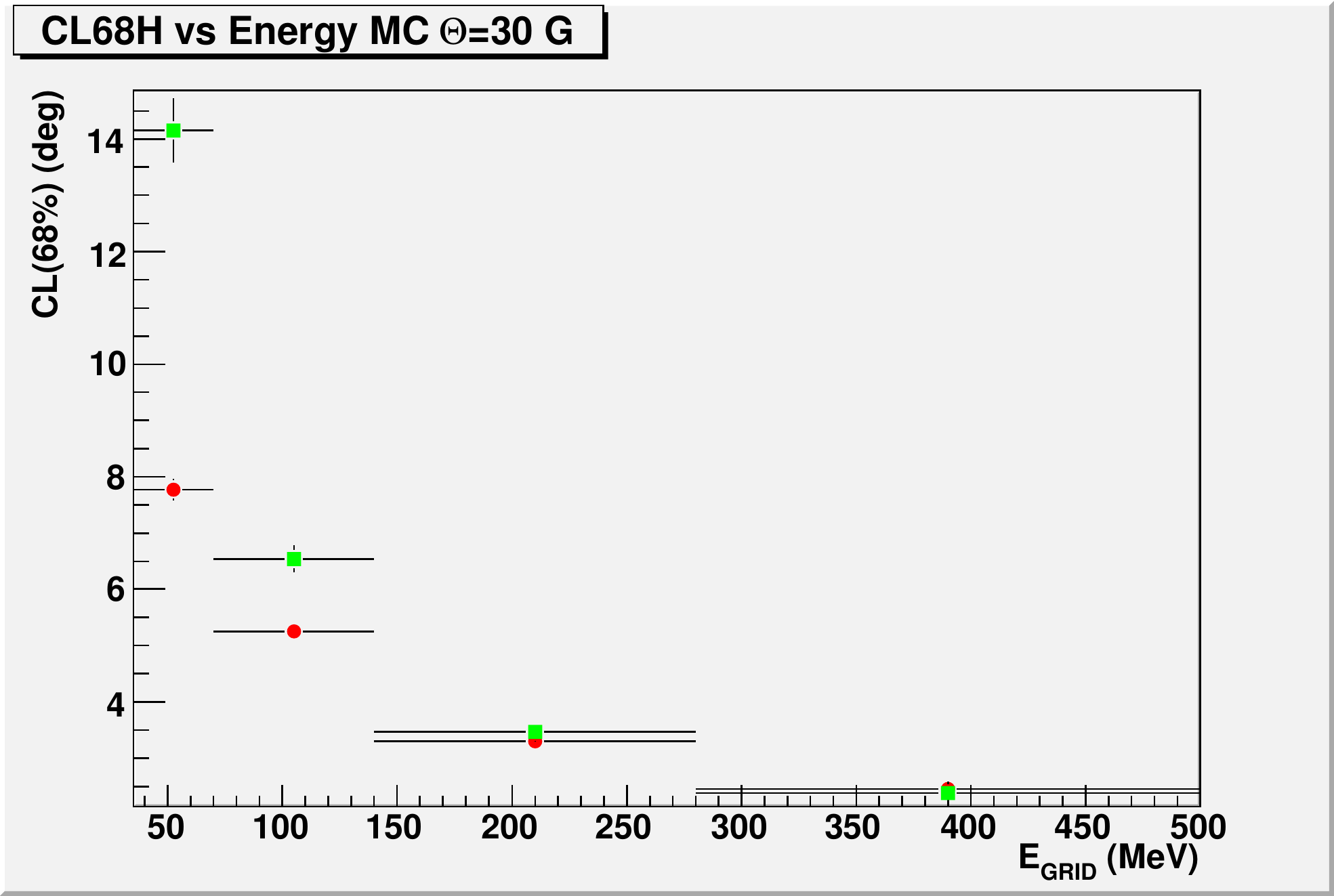}} &

\subfigure[]{\includegraphics[width=0.3\textwidth]{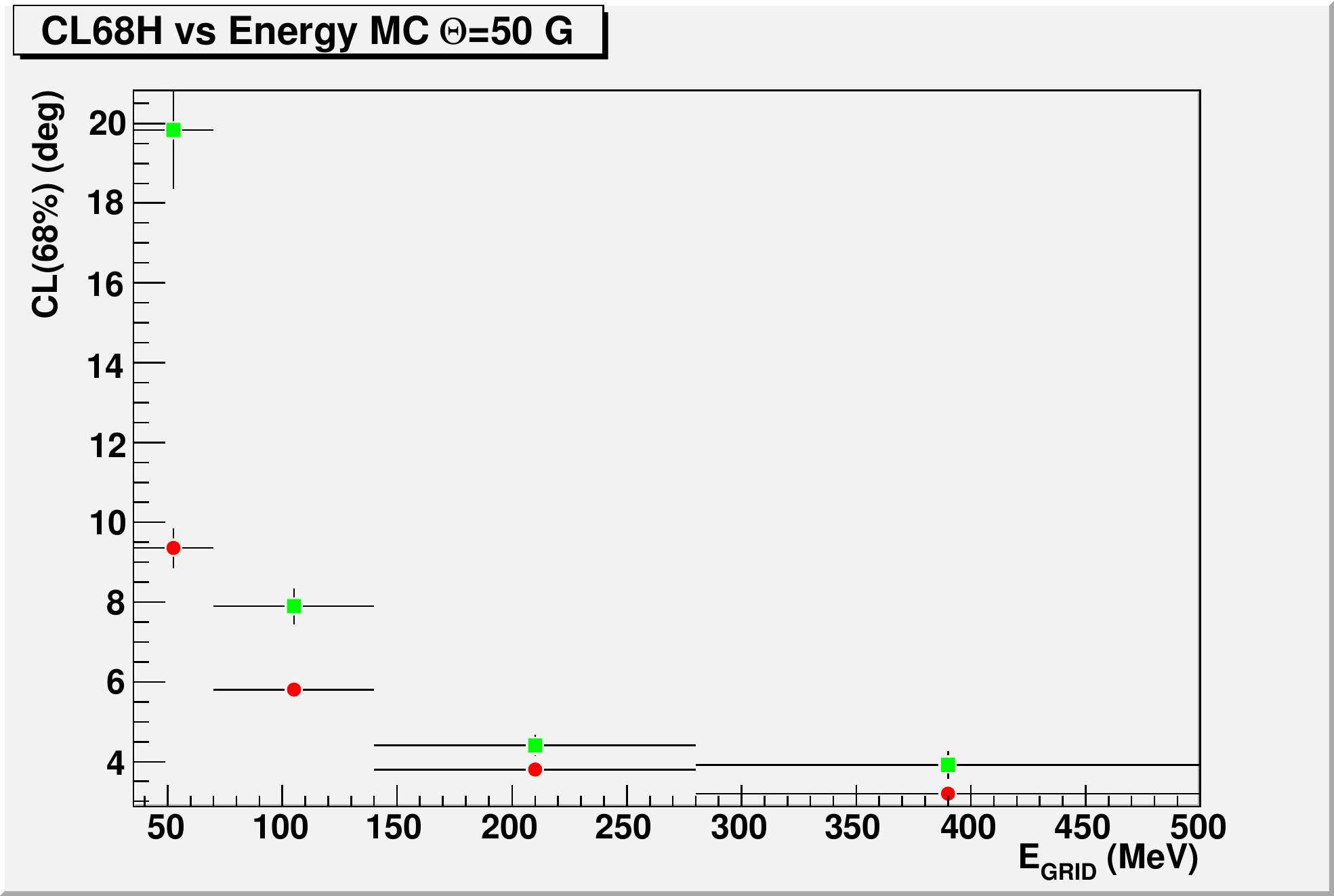}} 
\\
\end{tabular}}
\caption{Containment Radius 68.3\%\ of the King's function versus measured $\gamma$-ray energy $\egrid$ for real (green squares) and MC (red circles)
events a) $\Theta=\ang{0}$, b) $\Theta=\ang{30}$, c) $\Theta=\ang{50}$ 
}
\label{clemcreal}
\end{center}
\end{figure*}

The angular distributions are fitted
separately for different incident angles ($\Theta$),
different classes (G,L,S,P) and different intervals of energy. 
A subset of these fits is shown in Fig.\ref{psfkingmc} for MC events and
in Fig.\ref{psfkingreal} for real events for two bins of $\egrid$ for $\Theta=\ang{0}$ 
and $\Theta=\ang{30}$. The quality of all fits is good both for MC and real data.
In Fig.\ref{demcreal}-\ref{gemcreal}-\ref{clemcreal} the results of the King's function parameters 
obtained with the fits for MC and real data are shown versus $\egrid$.
In general the comparison is satisfying, although the parameter $\gamma$ for 
data is significantly lower than for MC especially for low $\egrid$. 
This implies larger Containment Radius boundaries for data at low energy.

\begin{figure*}[htb]
\begin{center}
\mbox{\begin{tabular}[t]{ccc}
\subfigure[]{\includegraphics[width=0.3\textwidth]{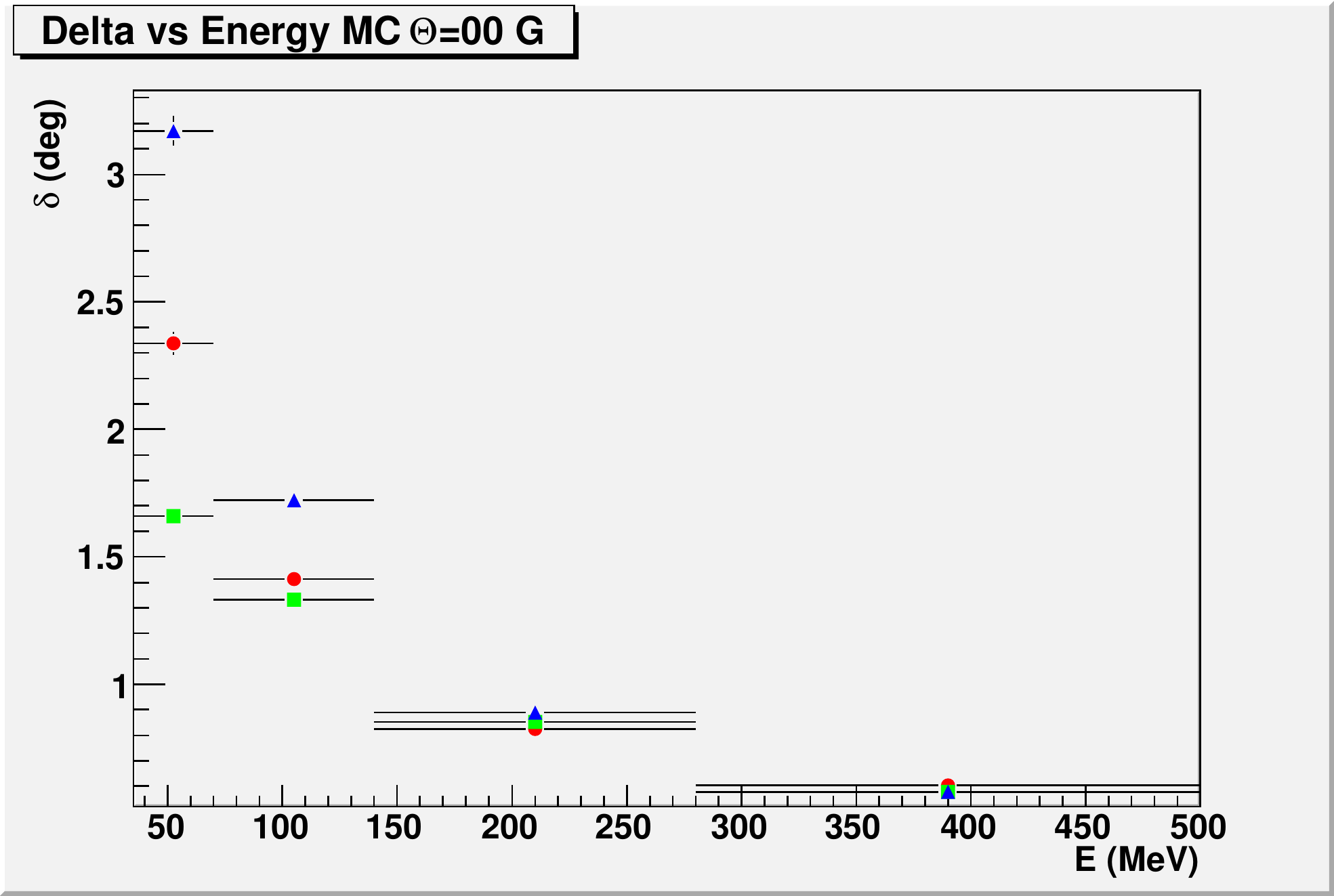}} &

\subfigure[]{\includegraphics[width=0.3\textwidth]{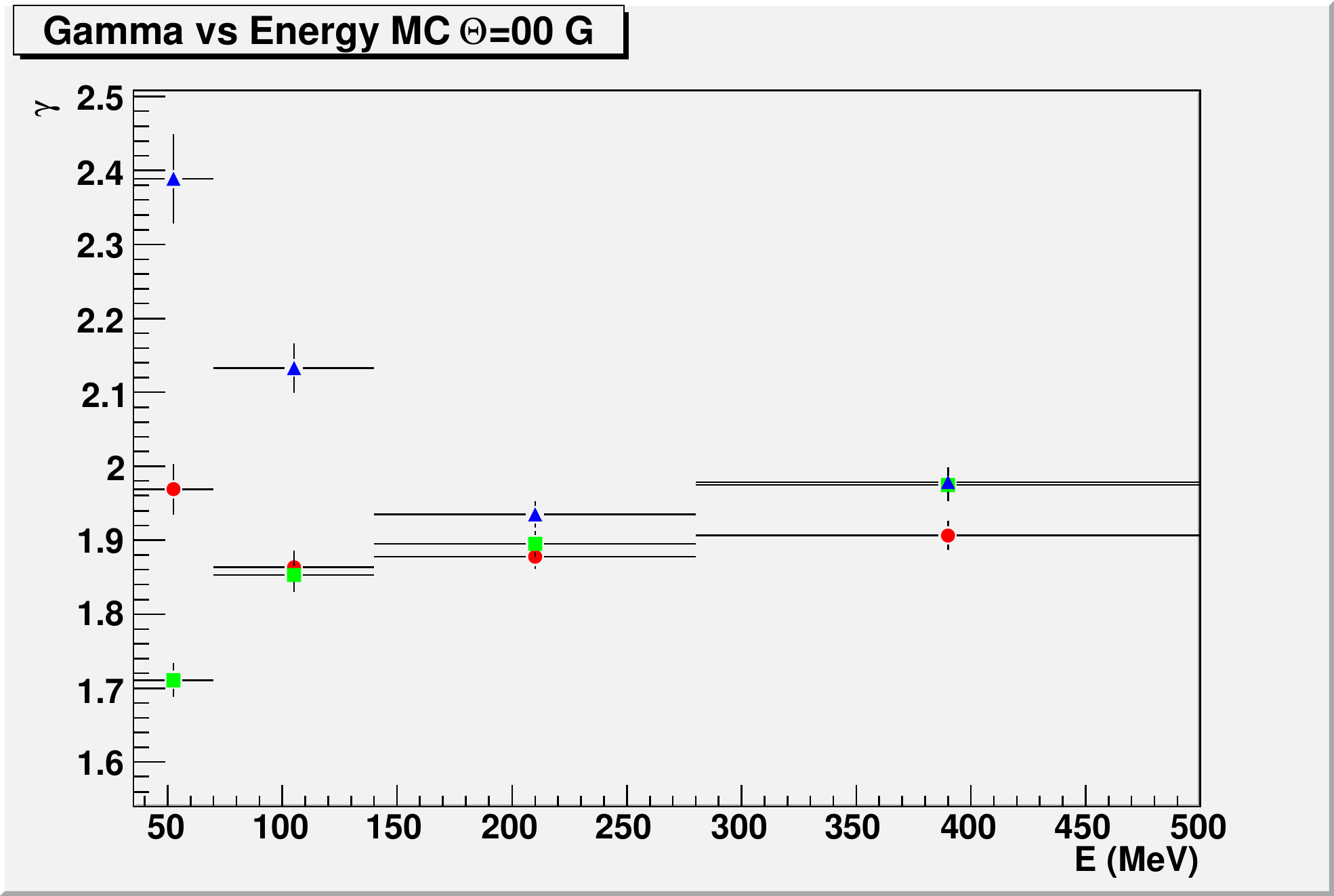}} &

\subfigure[]{\includegraphics[width=0.3\textwidth]{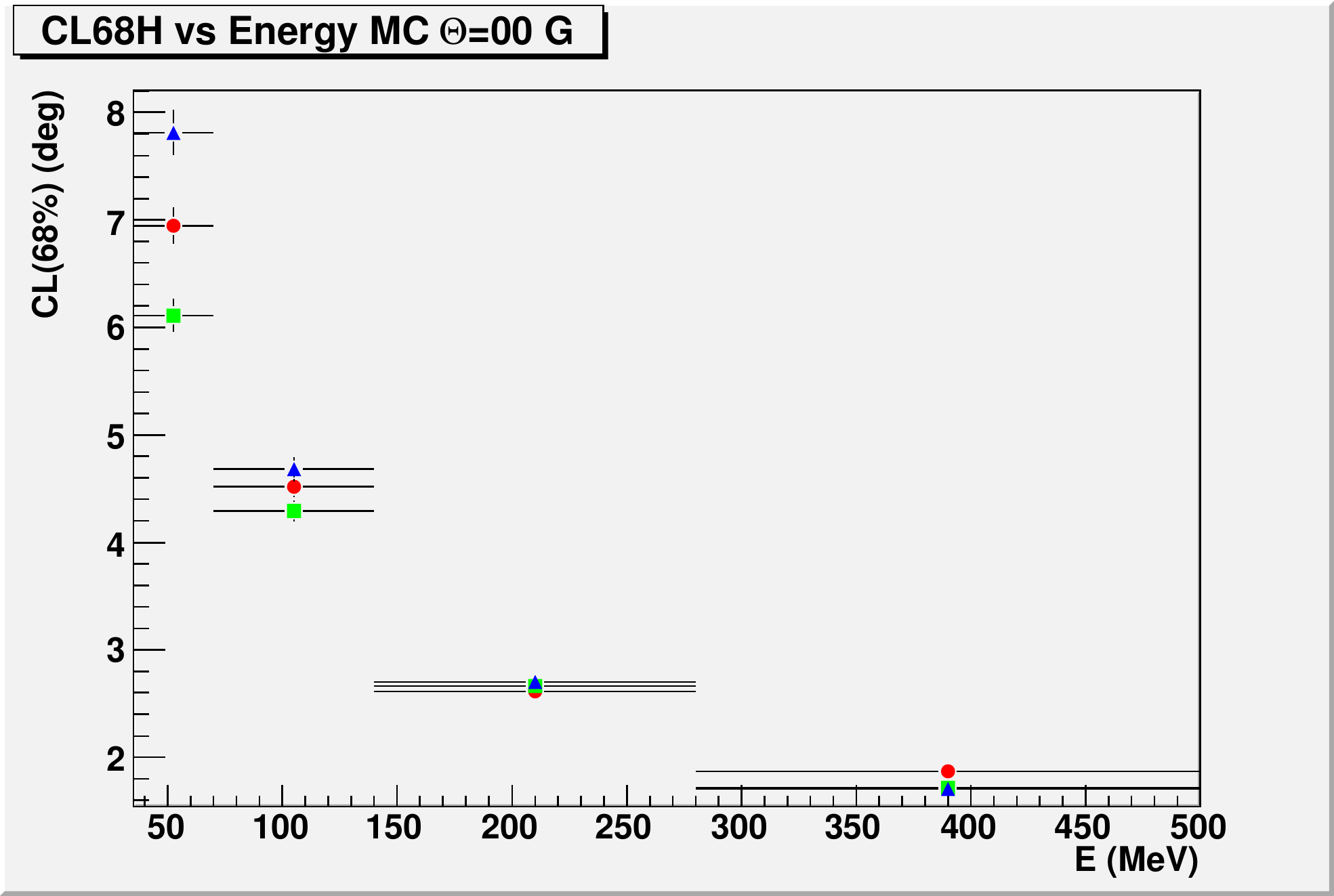}} 
\\
\end{tabular}}
\caption{
Parameters of the  King's function
for MC data at $\Theta$=\ang{0} versus different $\gamma$-ray energy estimator:  $\egrid$ (red circles), $\epts$ (green squares),
$\egamma$ (blue triangles). a) Parameter $\delta$ b) Parameter $\gamma$, c) PSF 
Containment Radius $CR_{68}$\%.
}
\label{kingmcenest}
\end{center}
\end{figure*}

\begin{figure*}[htb]
\begin{center}
\mbox{\begin{tabular}[t]{ccc}
\subfigure[]{\includegraphics[width=0.3\textwidth]{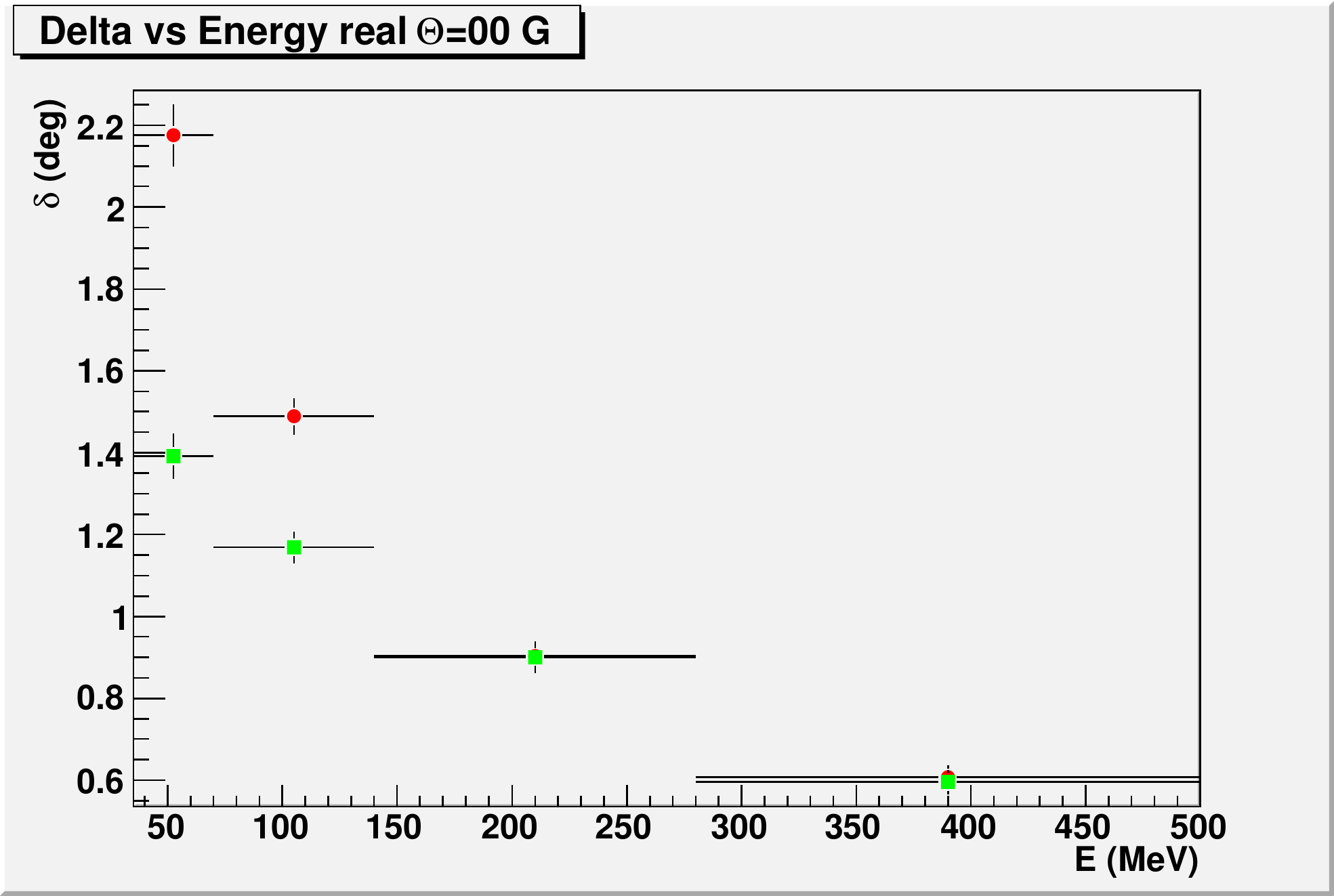}} &

\subfigure[]{\includegraphics[width=0.3\textwidth]{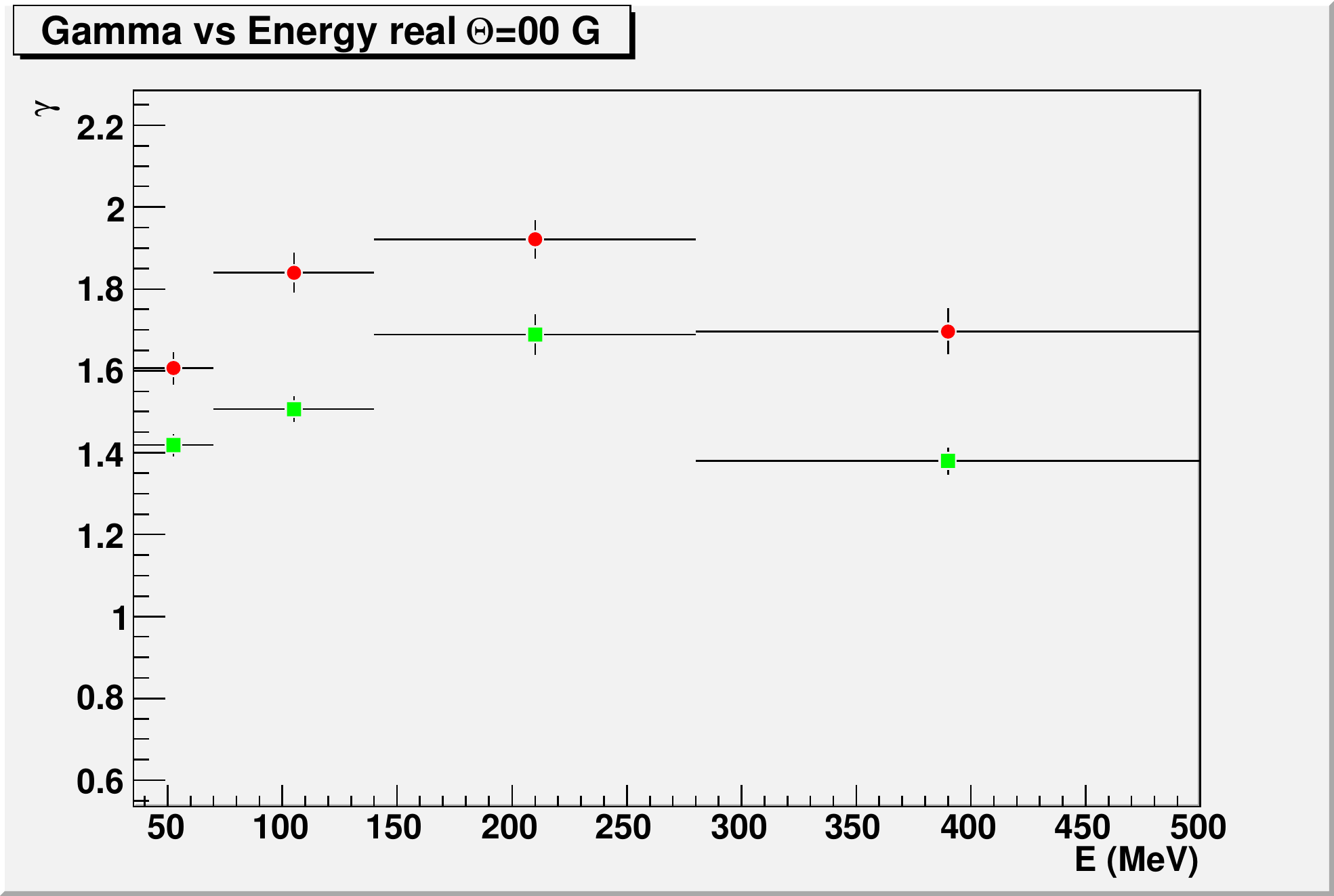}} &

\subfigure[]{\includegraphics[width=0.3\textwidth]{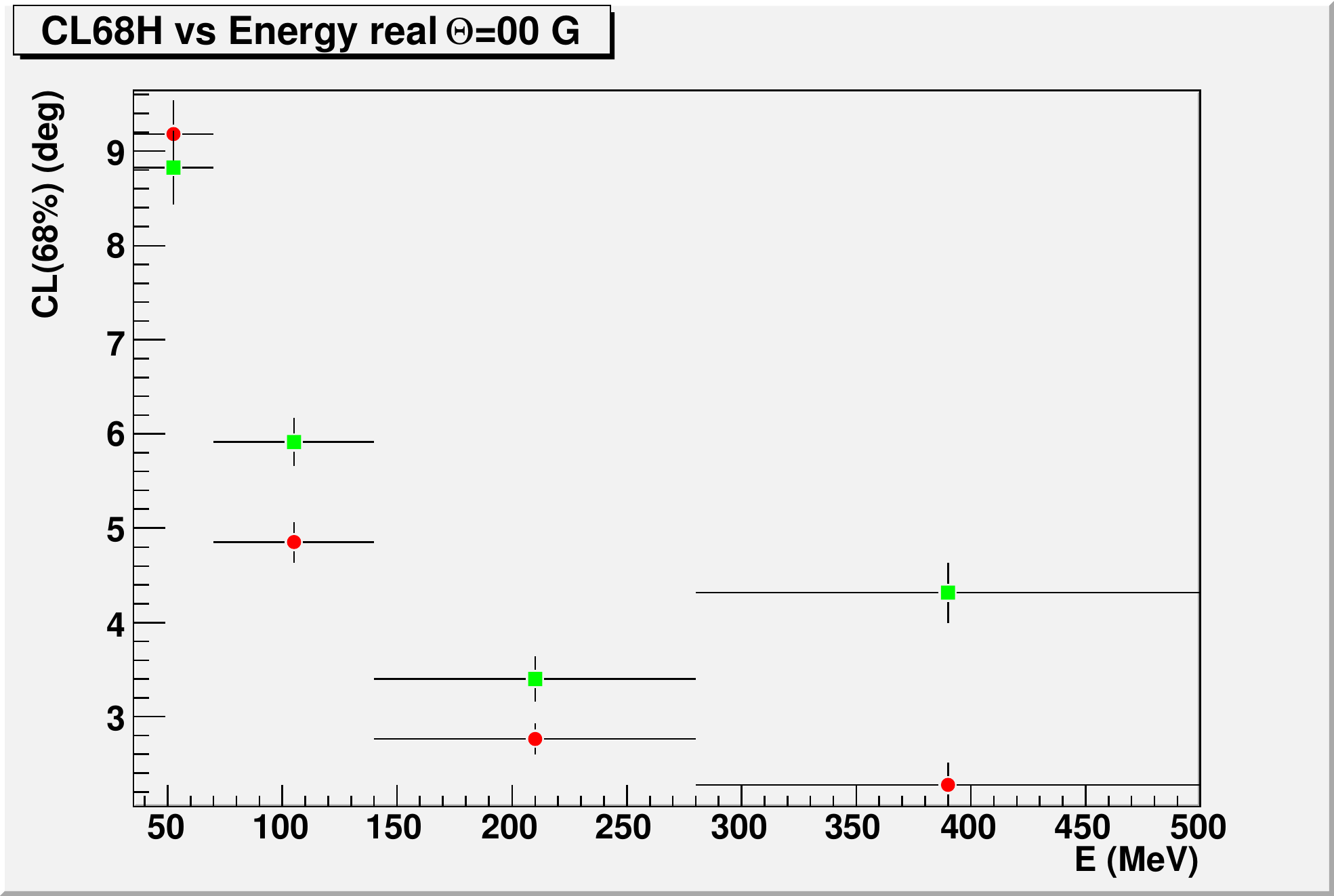}} 
\\
\end{tabular}}
\caption{
Parameters of the  King's function
for real data at $\Theta=\ang{0}$ for different $\gamma$-ray energy estimator:  $\egrid$ (red circles), $\epts$ (green squares).
a) $\delta$ b) $\gamma$, c) $CR_{68}$\%.
}
\label{kingrealenest}
\end{center}
\end{figure*}

Considering the broad EDP, especially at low energy, it is worth investigating if the choice of
the energy estimator influences significantly the PSF parameters. 
$\egrid$ and $\epts$ are available for both data and 
MC while only $\egamma$ is available for MC events.
In Fig.\ref{kingmcenest}-\ref{kingrealenest}, the parameters of the King's function from MC and real data
are displayed versus the relevant energy estimators. The agreement is good, the King's parameters for the 
same energy bin rarely differs more than 2-3 $\sigma$, 
with the exception of the last $\epts$ bin in real data, where the environmental 
background induces a large fraction of PSF events unrelated to real $\gamma$-rays of corresponding 
energy in the GRID as discussed in Sect.~\ref{ptsenerest}.

In Table~\ref{tabang} the results of the fits of the PSF distributions with the King's function are reported
versus the $\gamma$-ray energy for three incident angles for MC and real data.
The agreement is quite satisfactory with the exception of the parameter $\gamma$ and consequently $CR_{68}$ for
the lowest energy channel where real data PSFs show consistently longer tails.

\begin{table*}
\caption{King's function fit parameters of the PSF with the Containment Radius at 68.3\%\ for four energy bins and three 
angular direction for MC and real data.}
\label{tabang}
\begin{center}
\begin{tabular}{|c|c|c|c|c|c|c|c|}
\hline
$\Theta$&$\epts$ &\multicolumn{3}{c}{MC} \vline & \multicolumn{3}{c}{Real} \vline \\
\cline{3-8} 
(deg)&(\si{\MeV}) &$\delta$ (deg) & $\gamma$ & $CR_{68}$ (deg) &$\delta$ (deg) & $\gamma$ & $CR_{68}$ (deg) \\
\hline
0 &   35-70& 2.33$\pm$0.05 & 1.97$\pm$ 0.03 & 6.95$\pm$ 0.17 & 2.17$\pm$ 0.08 & 1.60$\pm$ 0.04 & 9.18$\pm$ 0.36 \\
  &  70-140& 1.41$\pm$0.02 & 1.86$\pm$ 0.02 & 4.52$\pm$ 0.10 & 1.49$\pm$ 0.04 & 1.84$\pm$ 0.05 & 4.85$\pm$ 0.22 \\
  & 140-180& 0.83$\pm$0.01 & 1.87$\pm$ 0.02 & 2.61$\pm$ 0.06 & 0.90$\pm$ 0.03 & 1.92$\pm$ 0.05 & 2.77$\pm$ 0.16 \\
  & 280-500& 0.60$\pm$0.01 & 1.91$\pm$ 0.02 & 1.87$\pm$ 0.06 & 0.61$\pm$ 0.03 & 1.69$\pm$ 0.06 & 2.27$\pm$ 0.23 \\
\hline
30 &   35-70& 2.72$\pm$0.05 & 2.03$\pm$ 0.04 & 7.77$\pm$ 0.19 & 2.42$\pm$ 0.09 & 1.45$\pm$ 0.03 &14.15$\pm$ 0.57 \\
   &  70-140& 1.62$\pm$0.02 & 1.85$\pm$ 0.02 & 5.25$\pm$ 0.10 & 1.48$\pm$ 0.04 & 1.58$\pm$ 0.03 & 6.54$\pm$ 0.25 \\
   & 140-180& 1.00$\pm$0.01 & 1.82$\pm$ 0.01 & 3.29$\pm$ 0.06 & 0.93$\pm$ 0.03 & 1.70$\pm$ 0.04 & 3.47$\pm$ 0.16 \\
   & 280-500& 0.70$\pm$0.01 & 1.76$\pm$ 0.01 & 2.45$\pm$ 0.05 & 0.69$\pm$ 0.03 & 1.77$\pm$ 0.05 & 2.38$\pm$ 0.22 \\
\hline
50 &   35-70& 2.67$\pm$0.12 & 1.76$\pm$ 0.08 & 9.36$\pm$ 0.17 & 3.09$\pm$ 0.22 & 1.41$\pm$ 0.07 & 19.83$\pm$ 1.47 \\
   &  70-140& 1.84$\pm$0.06 & 1.88$\pm$ 0.04 & 5.86$\pm$ 0.10 & 1.78$\pm$ 0.06 & 1.58$\pm$ 0.06 & 7.89$\pm$ 0.44 \\
   & 140-180& 1.05$\pm$0.02 & 1.76$\pm$ 0.02 & 3.86$\pm$ 0.06 & 1.04$\pm$ 0.04 & 1.60$\pm$ 0.04 & 4.40$\pm$ 0.26 \\
   & 280-500& 0.82$\pm$0.02 & 1.66$\pm$ 0.02 & 3.26$\pm$ 0.06 & 0.73$\pm$ 0.06 & 1.48$\pm$ 0.05 & 3.93$\pm$ 0.35 \\
\hline
\end{tabular}
\end{center}
\end{table*}

\subsection{Effective area \aeff}

This measurement turned out to be the most critical and we are unable to provide 
significant results. The reason is the strong flux of low energy photons in the 
experimental hall and, at a smaller extent, cosmic rays. 
This has been detected and discussed in depth in \citealt{btfagile},
that shows that many GRID triggers are fired outside the phase of the beam and some 
parts of the $\epts$ spectrum are unrelated to the $\gamma$-ray emission due to Bremsstrahlung.  
Furthermore even for events in phase, there are more hits on the front faces in real data 
that in MC.

The GRID detection efficiency is expected to be measured by the ratio of GRID events 
to PTS events in data and MC. The comparison can provide an estimation 
of the quality of the simulation and eventually provide correction factors.
This approach is sensible if the MC reproduces correctly the data and if the
calibration data resembles closely those expected in the space configuration.
The presence of a significant flux of cosmic rays triggering both the GRID and 
the PTS at random times are extremely difficult to simulate appropriately and 
outside the scope of the calibration task. 

An additional and even more relevant problem is the flux of low-energy photons 
in the experimental hall in phase with the beam. The low-energy photons down to 
X-ray energy are relevant because, if interacting with the silicon detectors,
they release enough energy to be above threshold for hit detection.
Those photons can be due to photon production 
along the beam line entering the hall, photon production in the last 
bending magnet, photon production in the beam dump of the electron beam followed by 
scattering inside the hall. This part is underestimated in the MC because it would 
require a simulation of the full hall for such low-energy photons that is unrealistic.
Such low-energy photon induced hits can overlap with $\gamma$-ray conversion events
forcing the filter to move e.g. an event from class G (clean conversion) to class L (ambiguous).

This effect is clearly visible in comparing MC (Fig.~\ref{psfkingmc})
and real data (Fig.~\ref{psfkingreal}) where the ratio between the numbers of G and L events 
is completely different being much lower for real data. Additional efforts
could bring the MC data closer to real data adding e.g. an additional flux of low-energy
photons hitting the GRID but it would be arbitrary and in any case 
significantly different from the space configuration so that little or no information
would be harvested.

Nevertheless this discrepancy is much less relevant for the measurements of the PSF and EDP
for class G events because for those measurements we rely on events
already classified as class G, that is for events for which the influence of the 
low-energy photons is non-existent or negligible. Under those conditions, the
comparison between MC and real data is realistic and the calibration results 
is expected to reproduce those in the space configuration.

\section{Conclusions}

In this paper the results of the calibration of the AGILE tracker GRID 
in the range \SIrange{35}{450}{\MeV}
on the BTF beam are presented.
In particular the EDPs and the PSFs versus the $\gamma$-ray energy and
the incident angle have been measured and compared with the MC expectations
with satisfactory results, while a reliable estimation of \aeff\ turned 
out to be unfeasible. 
From this comparison we quantify the 
reliability of the MC in describing the data and therefore in estimating
the systematic errors associated to the EDP matrices and PSF functions 
built with the MC and used in the analysis of in-flight AGILE data.
The differences in average values and in widths of the real and MC 
EDPs are mostly within 5\%\ as shown in Fig.\ref{psfener} (except for the highest energy).
The differences in PSFs mostly within 2-3 $\sigma$ with a few exceptions .

These results qualify AGILE as an instrument particularly effective in performing 
accurate measurements in the $\gamma$-ray energy range \SIrange{50}{450}{\MeV} as confirmed
by in-flight observations \citep{sabatini2010}.

\section*{Appendix}
\label{appen}

The PSF of instruments measuring high energy $\gamma$-rays is best characterised by the King's function
$k(\theta)$ \citep{king,chenaa}. 
The King's function can be defined by 
\begin{eqnarray}
\label{king}
dP(\theta) &=& k(\theta)\sin \theta d\theta  \nonumber \\
           &=& \frac{180}{\pi} \left(1-\frac{1}{\gamma}\right)\left(1+\frac{(\theta/\delta)^2}
   {2\gamma}\right)^{-\gamma} \frac{\sin\theta}{\delta} d\frac{\theta}{\delta}
\end{eqnarray}
where $\theta$ is the three-dimensional angular distance between the nominal and measured values
and $dP(\theta)$ is the probability of the angular distance to be between $\theta$ 
and $\theta + d\theta$.
The choice of the normalization factor is such that, under the approximation 
$\frac{180}{\pi} \sin\theta \approx \theta$, where $\theta$ is expressed in radian on the left
side and in degree on the right one, we obtain
\begin{eqnarray}
\label{kingint}
P(\theta) &=& \int_0^\theta k(\theta^\prime)\sin \theta^\prime d\theta^\prime = 
   1-\frac{1}{\left(1+\frac{(\theta/\delta)^2}{2\gamma}\right)^{\gamma-1}} \nonumber \\
P(+\infty) &=& 1
\end{eqnarray}

The standard deviation of the King's function is $\sigma = \delta\sqrt{\frac{\gamma}{\gamma-3/2}}$;
for $\gamma \to +\infty$ it converges to a Gaussian and $\sigma \to \delta$.
It is convenient to express the width of the angular PSF in terms of Containment Radius 
at 68.3\%\ ($CR_{68}$),
that is the angular value for which a predefined fraction of events (68.3\%\ in analogy with the 
fraction of events falling within a standard deviation for the Gaussian function) falls within the nominal direction.
If the PSF is parametrised by a King function, an approximate analytic calculation
of the $CR_{68}$ is possible.
In order to determine the value $CR$ such that $P(CR)=f$ ($0\le f\le 1$), the function in 
Eq.~\ref{kingint} can be inverted as
\begin{equation}
\label{kinginv}
CR_f = P^{-1}(f) = \delta\sqrt{2\gamma\left(\frac{1}{(1-f)^\frac{1}{\gamma-1}}-1\right)}
\end{equation}
For $\gamma\to +\infty$, The King's function becomes a Gaussian and Eq.\ref{kinginv} becomes 
\begin{eqnarray}
\label{gaussinv}
CR_f &=& \lim_{\gamma\to \infty} \delta\sqrt{2\gamma\left( \exp^{\frac{1}{\gamma-1}\ln\frac{1}{(1-f)}} - 1\right)} \nonumber \\
&=& \lim_{\gamma\to \infty} \delta\sqrt{{\frac{2\gamma}{\gamma-1}\ln{\frac{1}{(1-f)}}}} \nonumber \\
 &=& \delta \sqrt{2\ln\frac{1}{1-f}} \approx 1.5\, \delta \qquad \mathrm{for}\quad f=0.683
\end{eqnarray}

This result emphasises the difference between the CR of the PSF and that of the three-dimensional angular distribution.
Assuming a Gaussian PSF (and reporting as subscript the percentage rather than the fraction): 
for the PSF $CR_{68.3}= \delta$, $CR_{95.5}=2\delta$, $CR_{99.7}=3\delta$; 
for the three-dimensional distribution, $CR_{68.3}\approx 1.51\,\delta$, $CR_{95.5}\approx 2.49\,\delta$, 
$CR_{99.7}\approx 3.44\,\delta$.

\section*{Acknowledgments}

The authors would like to thank the staff of the BTF at Laboratori 
Nazionali di Frascati who made this work possible.

\bibliographystyle{aa}
\bibliography{GRIDPaper}

\end
{document}